\documentclass[a4paper,10pt,twoside]{cpc-hepnp}

\usepackage[T1,T2A]{fontenc}
\usepackage[utf8]{inputenc}
\inputencoding{cp1251}
\usepackage{multicol}
\usepackage{graphicx}
\usepackage{booktabs}
\usepackage{amssymb,bm,mathrsfs,bbm,amscd}
\usepackage{verbatim}
\usepackage[tbtags]{amsmath}
\usepackage{lastpage}
\usepackage{CJK}

\begin{document}
%\begin{CJK*}{GBK}{song}

\fancyhead[c]{\small Yu.G. Ignat'ev, A.A. Agathonov, I.A. Kokh}
\fancyfoot[C]{\small 010201-\thepage}

\footnotetext[0]{Received 23 October 2018}

\title{The Peculiarities of the Cosmological Models Based on Nonlinear Classical and Phantom Fields with Minimal Interaction. II. The Cosmological Model Based on the Asymmetrical Scalar Doublet\thanks{The work is performed according to the Russian Government Program of Competitive Growth of Kazan Federal University}}

\author{%
      Yurii Ignat'ev$^{1)}$\email{ignatev-yurii@mail.ru}%
\quad Alexander Agathonov$^{2)}$\email{a.a.agathonov@gmail.com}%
\quad Irina Kokh$^{2)}$\email{irina\_kokh@rambler.ru}
}
\maketitle

\address{%
$^1$ Institute of Physics, Kazan Federal University, $^2$ Lobachevsky Institute of Mathematics and Mechanics, Kazan Federal University,  Kremleovskay str. 18, Kazan,  420008,  Russia\\
}

\begin{abstract}
A detailed comparative qualitative analysis and numerical simulation of evolution of the cosmological models based on the doublet of classical and phantom scalar fields with self-action. The 2-dimensional and 3-dimensional projections of the phase portraits of the corresponding dynamic system are built. Just as in the case of single scalar fields, the phase space of such systems, becomes multiply connected, the ranges of negative total effective energy unavailable for motion, getting appear there. The distinctive feature of the asymmetrical scalar doublet is the time dependency of the prohibited ranges' projections on the phase subspaces of each field as a result of which the existence of the limit cycles with null effective energy depends on the parameters of the field model and initial conditions. The numerical models where the dynamic system has limit cycles on hypersurfaces of null energy, are built. It is shown, that even quite weak phantom field in such model undertakes functions of management of the dynamic system and can significantly change the course of the cosmological evolution.
\end{abstract}



\footnotetext[0]{\hspace*{-3mm}\raisebox{0.3ex}{$\scriptstyle\copyright$}2013
Chinese Physical Society and the Institute of High Energy Physics
of the Chinese Academy of Sciences and the Institute
of Modern Physics of the Chinese Academy of Sciences and IOP Publishing Ltd}%

\begin{multicols}{2}

%%%%%%%%%%%%%%%%%%%%%%%%%%%%%%%%%%%%%%%%%%%%%%%%%%%%%%%%%%%%%%%%%%%%%%%%%%%%%%%%%%%%%%%%%%%%%%%%%%%%%
\tableofcontents
\section{The Basic Relations of the Cosmological Model Based on the Asymmetrical Scalar Doublet}
\subsection{The Lagrangian Function and the Potential of Self-Action}
In the previous paper of the Author \cite{Part1} the cosmological models based on single classical and phantom scalar fields with self-action were investigated. Let us now consider a system comprising two minimally interacting scalar fields, classical and phantom ones, which we will call
\emph{an asymmetrical scalar doublet}. The Lagrangian function of the scalar doublet comprising classical and phantom scalar fields with self-action in the Higgs form with minimal coupling has the following form \cite{Yu_STFI_3,YuIrina_Phys}:
\begin{equation} \label{Eq__1_}
L=\frac{1}{8\pi } (g^{ik} \Phi _{,i} \Phi _{,k} -2V(\Phi ))-\frac{1}{8\pi } (g^{ik} \varphi _{,i} \varphi _{,k} +2v(\varphi )),
\end{equation}
where
\begin{eqnarray}
\label{V}
V(\Phi )=-\frac{\alpha }{4} \left(\Phi ^{2} -e\frac{m^{2} }{\alpha } \right)^{2} ;\\
\label{v}
v(\varphi )=-\frac{\beta }{4} \left(\varphi ^{2} -\varepsilon \frac{\mathfrak{m}^{2} }{\beta } \right)^{2}
\end{eqnarray}
-- the potential energy of the corresponding scalar fields, $\alpha$ and $\beta$ are the constants of their self-action, $m$ and $\mathfrak{m}$ are their quanta masses. Introducing summary potential

\begin{eqnarray} \label{Eq__2_}
U(\Phi ,\varphi )=V(\Phi )+v(\varphi )=-\frac{\alpha }{4} \left(\Phi ^{2} -e\frac{m^{2} }{\alpha } \right)^{2} \nonumber\\
-\frac{\beta }{4} \left(\varphi ^{2} -\varepsilon \frac{\mathfrak{m}^{2} }{\beta } \right)^{2} \equiv U(e,\varepsilon ,\alpha ,\beta ;\Phi ,\varphi ),
\end{eqnarray}
we can make the following conclusions:

\noindent 1. The potential $U(\Phi ,\varphi )$ possess the following symmetries:

\begin{equation} \label{Eq__3_}
U(\pm \Phi ,\pm \varphi )=U(\Phi ,\varphi );
\end{equation}

\begin{equation} \label{Eq__4_}
U(-e,-\varepsilon ,-\alpha ,-\beta ;\Phi ,\varphi )=-U(e,\varepsilon ,\alpha ,\beta ;\Phi ,\varphi ).
\end{equation}
2. The function $U(\Phi ,\varphi )$ at $\{ e=1,\varepsilon =1\} $ has the absolute maximum in the origin of coordinates of the phase space $\{ \Phi ,\varphi \} $ $M_{0} (0,0)$, and at $\{ e=-1,\varepsilon =-1\} $ it has the absolute minimum: $e\varepsilon =1$ everywhere, it has a conditional extremum in these points (saddle points) at $e\varepsilon =-1$.

\noindent 3. The function $U(\Phi ,\varphi )$  has its absolute maximum at  $e\alpha >0$ and $\varepsilon \alpha <0$ in $M_{10} (-m/\sqrt{e\alpha } ,0)$ and $M_{20} (m/\sqrt{e\alpha } ,0)$ at $\alpha <0$ (i.e. $\alpha <0,e=-1,\varepsilon =+1$) and the absolute minimum in these points at $\alpha >0$ (i.e. $\alpha >0,e=+1,\varepsilon =-1$):  $e\varepsilon =-1$ everywhere; it has a conditional extremum in these points (saddle points) at $e\alpha >0$ and $\varepsilon \alpha >0$: $e\varepsilon =1$ everywhere.

\noindent 4. The function $U(\Phi ,\varphi )$ has its absolute maximum at $\varepsilon \beta >0$ and $e\beta <0$ in the points $M_{01} 
(0,-m/\sqrt{\varepsilon \beta } )$ and $M_{02} (0,m/\sqrt{\varepsilon \beta } $ at $\beta <0$ (i.e. $\beta <0,e=+1,\varepsilon =-1$) and the 
absolute minimum in these points at $\beta >0$ (i.e. $\beta >0,e=-1,\varepsilon =+1$): $e\varepsilon =-1$ everywhere; it has a conditional 
extremum in these points (saddle points) at $\varepsilon \beta >0$ and $e\beta >0$ , i.e., at $e\varepsilon =1$.

\noindent 5. The function $U(\Phi ,\varphi )$ has its absolute maximum at  $e\alpha >0$,$\varepsilon \beta >0$ and $\alpha \beta >0$ in the points $M_{11} (-m/\sqrt{e\alpha } ,-m/\sqrt{\varepsilon \beta } )$, $M_{12} (-m/\sqrt{e\alpha } ,m/\sqrt{\varepsilon \beta } )$, $M_{21} (m/\sqrt{e\alpha } ,-m/\sqrt{\varepsilon \beta } )$ and $M_{22} (m/\sqrt{e\alpha } ,m/\sqrt{\varepsilon \beta } )$ at $\alpha >0$ (i.e. $\alpha >0,\beta >0,e=1,\varepsilon =1$) and the absolute minimum in these points at $\alpha <0$: $e\varepsilon =1$ everywhere; it has a conditional extremum in these points (saddle points) at $\alpha \beta <0$, i.e. at $e\varepsilon =-1$.
\end{multicols}
\ruleup
\begin{center}
\includegraphics[width=16cm]{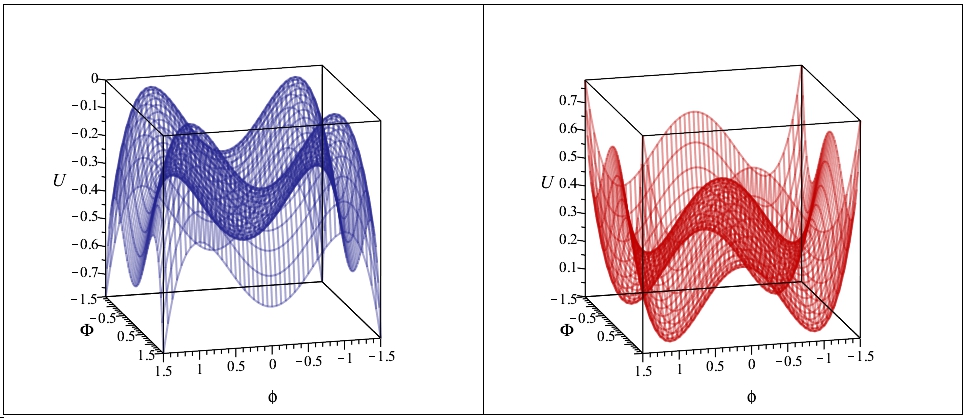}
\figcaption{\label{ris1} On the left-hand side: the graph of the potential $U(1,1,1,1;\Phi ,\varphi )$, on the right-hand side: the graph of the potential $U(-1,- 1,- 1,- 1;\Phi,\varphi )$ }
\end{center}
\ruledown

\begin{multicols}{2}

\noindent The typical graphs of the potential function $U(\Phi ,\varphi )$, corresponding to two opposite cases described in \eqref{Eq__5_} are shown on Fig. \ref{ris1}. On the left graph one can see 4 maximums, 1 central minimum and 4 saddle points while on the right one -- 4 minimums and one central maximum as well as 4 saddle points. The right figure is obviously obtained by mirroring of the left one from the plane $U=0$.

 Thus, taking into account the fact that stationary points of the dynamic system with the Lagrangian function of the form \eqref{Eq__1_} coincide with the stationary points of the potential $U(\Phi ,\varphi )$, we can state the following: depending on the signs of the parameters $\{ e,\varepsilon ,\alpha ,\beta \} $ potential $U(e,\varepsilon ,\alpha ,\beta ;\Phi ,\varphi )$, the corresponding dynamic system should have 1, 3 or 9 stationary points among which there should be the attractive (absolute minimum), repulsing  (absolute maximum) and saddle  (conditional extremum) points. This result fully coincides with the conclusions of the paper \cite{Yu_STFI_3}, where it was obtained with the help of qualitative theory of the dynamic systems.

\subsection{The Equations of the Cosmological Model}
The energy-momentum tensor of the scalar field relative to the Lagrangian function \eqref{Eq__1_} takes the standard form:
\begin{eqnarray} \label{Eq__5_}
T_{ik} =\frac{1}{8\pi } (2\Phi _{,i} \Phi _{,k} -g_{ik} \Phi _{,j} \Phi ^{,j} +2V(\Phi )g_{ik} )\nonumber\\
-\frac{1}{8\pi } (2\varphi _{,i} \varphi _{,k} +g_{ik} \varphi _{,j} \varphi ^{,j} -2v(\varphi )g_{ik} ).
\end{eqnarray}
The Lagrangian function's variation \eqref{Eq__1_} leads to the next field equations:
\[\begin{array}{l} {\square \Phi +V'(\Phi )=0;{\rm \; }} \\ {\square \varphi +v'(\varphi )=0.} \end{array}\]
Carrying out renormalization of the Lagrangian function \eqref{Eq__1_}, and adding a constant to it, (see [2]), let us reduce it to the following form:
\begin{eqnarray} \label{Eq__6_}
L=\frac{{\rm 1}}{8\pi } \left(g^{ik} \Phi _{,i} \Phi _{,k} -em^{2} \Phi ^{2} +\frac{\alpha }{2} \Phi ^{4} \right)\nonumber\\
-\frac{{\rm 1}}{8\pi } \left(g^{ik} \varphi _{,i} \varphi _{,k} +\varepsilon \mathfrak{m}^{2} \varphi ^{2} -\frac{\beta }{2} \varphi ^{4} \right).
\end{eqnarray}
The corresponding renormalzation of the energy-momentum tensor gives us
\begin{equation} \label{Eq__7_}
\begin{array}{c} {T_{ik} =\displaystyle\frac{1}{8\pi } \left(2\Phi _{,i} \Phi _{,k} -g_{ik} \Phi _{,j} \Phi ^{,j} +g_{ik} em^{2} \Phi ^{2} -g_{ik} \frac{\alpha }{2} \Phi ^{4} \right)} \\ {-\frac{{\rm 1}}{8\pi } \left(2\varphi _{,i} \varphi _{,k} -g_{ik} \varphi _{,j} \varphi ^{,j} -g_{ik} \varepsilon \mathfrak{m}^{2} \varphi ^{2} +g_{ik} \frac{\beta }{2} \varphi ^{4} \right).} \end{array}
\end{equation}
The equations of free classical and phantom fields can be found by means of the standard variational procedure over the Lagrangian function in the form \eqref{Eq__6_}:
\begin{eqnarray} \label{Eq__8_}
\square \Phi +m_{*}^{2} \Phi =0;\\
\label{Eq__9_}
\square \varphi +\mathfrak{m}_{*}^{2} \varphi =0,
\end{eqnarray}
where $m_{*}^{} $ and $\mathfrak{m}_{*}^{} $ are effective masses of scalar bosons:
\begin{equation} \label{Eq__10_}
\begin{array}{l} {m_{*} ^{2} =em^{2} -\alpha \Phi ^{2} ;} \\ {\mathfrak{m}_{*} ^{2} =-\varepsilon \mathfrak{m}^{2} +\beta \varphi ^{2} .} \end{array}
\end{equation}
Let us further consider a self-consistent system of equations of the cosmological model \eqref{Eq__8_}, \eqref{Eq__9_} and Einstein equations \eqref{EqEinstein0}
\footnote{Here we use the Planck system of units: $G=c=\hbar =1$; the Ricci tensor is obtained by means of convolution of the first and fourth indices $R_{ik}=R^j_{~ikj}$; the metrics has the signature $(-1,-1,-1,+1)$.}
\begin{equation} \label{EqEinstein0}
R^{ik} -\frac{1}{2} Rg^{ik} =\lambda g^{ik} +8\pi T^{ik},
\end{equation}
where $\lambda\geq 0$ is the cosmological constant based on free assymetrical scalar doublet and space-flat Friedmann metrics (\ref{metric})
\begin{equation}\label{metric}
ds^{2} =dt^{2} -a^{2} (t)(dx^{2} +dy^{2} +dz^{2}),
\end{equation}
assuming
$\Phi =\Phi (t),$ $\varphi =\varphi (t).$ The energy-momentum tensor at that \eqref{Eq__7_} takes a structure of the energy-momentum tensor of the isotropic liquid with energy density $\mathcal{E}$ and pressure \textit{p}:
\begin{eqnarray}
\label{Eq__13_}
\mathcal{E}(t)=\mathcal{E}_{c} +\mathcal{E}_{f} ,{\rm \; \; }\, \, p=p_{c} +p_{f} ;\\
 \label{Eq__14_}
\mathcal{E}_{c} =\frac{1}{8\pi } \left(\dot{\Phi }^{2} +em^{2} \Phi ^{2} -\frac{\alpha }{2} \Phi ^{4} \right);\nonumber\\
\mathcal{E}_{f} =\frac{1}{8\pi } \left(-\dot{\varphi }^{2} +\varepsilon \mathfrak{m}^{2} \varphi ^{2} -\frac{\beta }{2} \varphi ^{4} \right);\nonumber\\
p_{c} =\frac{1}{8\pi } \left(\dot{\Phi }^{2} -em^{2} \Phi ^{2} +\frac{\alpha }{2} \Phi ^{4} \right);\\
p_{f} =\frac{1}{8\pi } \left(-\dot{\varphi }^{2} -\varepsilon \mathfrak{m}^{2} \varphi ^{2} +\frac{\beta }{2} \varphi ^{4} \right)\nonumber.
\end{eqnarray}
Herewith the followind identity law is fulfilled:
\begin{equation} \label{Eq__15_}
\mathcal{E}+p\equiv \frac{1}{4\pi } \dot{\Phi }^{2} -\frac{1}{4\pi } \dot{\varphi }^{2} .
\end{equation}

The considered system comprises one Einstein equation
\begin{eqnarray} \label{Eq__16_}
3\frac{\dot{a}^{2} }{a^{2} } \equiv 3H^{2} =\left(\dot{\Phi }^{2} +em^{2} \Phi ^{2} -\frac{\alpha }{2} \Phi ^{4} \right)-\nonumber\\
\left(\dot{\varphi }^{2} -\varepsilon \mathfrak{m}^{2} \varphi ^{2} +\frac{\beta }{2} \varphi ^{4} \right)+\lambda
\end{eqnarray}
and two equations of the scalar field:
\begin{equation} \label{Eq__17_}
\ddot{\Phi }+3\frac{a}{a} \dot{\Phi }+m_{*}^{2} \Phi =0,
\end{equation}
\begin{equation} \label{Eq__18_}
\ddot{\varphi }+3\frac{\dot{a}}{a} \dot{\varphi }+\mathfrak{m}_{*}^{2} \varphi =0.
\end{equation}

Substituting the expressions for the effective masses $m_{*}^{} $ and $\mathfrak{m}_{*}^{} $ \eqref{Eq__10_} into \eqref{Eq__14_}, \eqref{Eq__15_} we find a final form  of the system of equations:
\begin{eqnarray} \label{Eq__19_}
3\frac{\dot{a}^{2} }{a^{2} } =\bigl(\dot{\Phi }^{2} +em^{2} \Phi ^{2} -\frac{\alpha }{2} \Phi ^{4} \bigr)-\nonumber\\
\bigl(\dot{\varphi }^{2} -\varepsilon \mathfrak{m}^{2} \varphi ^{2} +\frac{\beta }{2} \varphi ^{4} \bigr)+\lambda ;\\
\label{Eq__20_}
\ddot{\Phi }+3\frac{\dot{a}}{a} \dot{\Phi }+em_{}^{2} \Phi -\alpha \Phi ^{3} =0;\\
\label{Eq__21_}
\ddot{\varphi }+3\frac{\dot{a}}{a} \dot{\varphi }-\varepsilon \mathfrak{m}_{}^{2} \varphi +\beta \varphi ^{3} =0.
\end{eqnarray}

\section{Qualitative Analysis of the Cosmological Model}

\subsection{Reducing the System of Equations to the Canonical Form}

Proceeding to the dimensionless Compton time: $mt=\tau ;$ ($m\ne 0$) and carrying out a standard change of variables:
\begin{equation} \label{Eq__23_}
\Phi '=Z(\tau ),\, \, \, \, \, \varphi '=z(\tau ),\, \, \, \, \, (f'\equiv df{\rm /}d\tau ),
\end{equation}
introducing further according to \eqref{V} and \eqref{v} the potential energy of scalar $V(\Phi )$, and phantom, $v(\varphi )$ fields let us write down the expression for the \emph{effective reduced energy density}
\begin{eqnarray} \label{E_m}
\mathcal{E}_m(\Phi,Z,\varphi,z) =\mathcal{E}_{c} +\mathcal{E}_{f} +\Lambda_m =\nonumber\\
\left(\frac{Z^{2} }{2} -V(\Phi )\right)-\left(\frac{z^{2} }{2} -v(\varphi )\right)+\Lambda_m ,
\end{eqnarray}
where the total energies for classical and phantom fields are introduced:
\begin{eqnarray} \label{Eqs7_}
\mathcal{E}_{c} =\frac{Z^{2} }{2} -V(\Phi );\quad \mathcal{E}_f =-\left(\frac{z^{2} }{2} -v(\varphi )\right),\\
\alpha_m=\frac{\alpha}{m^2},\quad \beta_m=\frac{\beta}{m^2},\quad \lambda_m=\frac{\lambda}{m^2}.\nonumber\\
\label{Eqs6_}
\Lambda _{m} =\lambda _{m} -\frac{1}{2\alpha _{m} } -\frac{\mu ^{2} }{2\beta _{m} },\quad \mu\equiv \frac{\mathfrak{m}}{m}
\end{eqnarray}
Let us also introduce the effective reduced pressure of the scalar doublet:
\begin{eqnarray} \label{p_m}
p_m(\Phi,Z,\varphi,z) =p_{c} +p_{f} -\Lambda_m =\nonumber\\
\left(\frac{Z^{2} }{2} +V(\Phi )\right)-\left(\frac{z^{2} }{2} +v(\varphi )\right)-\Lambda_m ,
\end{eqnarray}
so that it is
\begin{equation}\label{e+p}
\mathcal{E}_m(\Phi,Z,\varphi,z)+p=Z^2-z^2.
\end{equation}
This, let us reduce the Einstein equation \eqref{Eq__19_} to the dimensionless form:
\begin{eqnarray} \label{Eq__24_}
\frac{{a'}^{2} }{a^{2} }\equiv {H'}_{m} ^{2}&\displaystyle =\frac{1}{3} \mathcal{E}_m(\Phi,Z,\varphi,z),
\end{eqnarray}
and the field equations \eqref{Eq__20_}, \eqref{Eq__21_} -- to the form of normal autonomous system of ordinary differential equations in a 4-dimensional phase space ${\mathbb R}_{4} :\{ \Phi ,Z,\varphi ,z\} $
\begin{eqnarray} \label{Dyn_sys}
 \Phi '=Z; & Z'= - Z\sqrt{3\mathcal{E}_{m}(\Phi,Z,\varphi,z)} -e\Phi +\alpha _{m} \Phi ^{3} ; \\[12pt]
 \varphi '=z;&  z'= -z\sqrt{3\mathcal{E}_{m}(\Phi,Z,\varphi,z)}  +\varepsilon \mu ^{2} \varphi -\beta _{m} \varphi ^{3} .\nonumber
\end{eqnarray}
Beforehand, let us notice that the conclusions of the qualitative theory relative to stationary points of the dynamic system and their character can't differ from the conclusions obtained on the basis of the potential function's analysis. However, the qualitative theory provides more details of the dynamic system's behavior in the neighbourhood of stationary points.

\noindent In order the system of differential equations \eqref{Dyn_sys} to have a real solution, the non-negativeness of the expression beneath the radical in the equations i.e. non-negativeness of the system's effective energy with an account of the cosmological term, is required:
\begin{eqnarray} \label{Eq__26_}
\mathcal{E}_{m}(\Phi,Z,\varphi,z)\equiv Z^{2} +e\Phi ^{2} -\frac{\alpha _{m} }{2} \Phi ^{4}\nonumber\\
 -z^{2} +\varepsilon \mu ^{2} \varphi ^{2} -\frac{\beta _{m} }{2} \varphi ^{4}+\lambda _{m} \ge 0.
\end{eqnarray}
The inequality \eqref{Eq__26_} can lead to breaking of the simple connectedness of the phase space and generation of closed lacunae in it, which are limited by surfaces with null effective energy. To reduce the system \eqref{Dyn_sys} to the standard notation of the qualitative theory (see, e.g. \cite{Bautin})
\begin{equation}\label{sys_norm}
\frac{dx_{i} }{d\tau } =F_{i} (x_{1} ,\ldots ,x_{n} ),{\rm }i=\overline{1,n}
\end{equation}
let us accept the following denotations:

\end{multicols}
\ruleup
\begin{equation} \label{Eq__27_}
\begin{array}{l} {\Phi =x\quad (=x_1);\quad\varphi =y\quad (=x_2);\quad F_{1} \equiv P=Z\quad (=x_3);\quad F_{3} \equiv p=z\quad (=x_4);} \\
F_{2} \equiv Q=-\sqrt{3} Z\sqrt{\displaystyle\left(Z^{2} +ex^{2} -\frac{\alpha _{m} }{2} x^{4} \right)-\left(z^{2} -\varepsilon \mu ^{2} y^{2} +\frac{\beta _{m} }{2} y^{4} \right)+\lambda _{m} } -ex+\alpha _{m} x^{3} ; \\
F_{4} \equiv q=-\sqrt{3} z\sqrt{\displaystyle\left(Z^{2} +ex^{2} -\frac{\alpha _{m} }{2} x^{4} \right)-\left(z^{2} -\varepsilon \mu ^{2} y^{2} +\frac{\beta _{m} }{2} y^{4} \right)+\lambda _{m} } +\varepsilon \mu ^{2} y-\beta _{m} y^{3} . \end{array}
\end{equation}
\ruledown

\begin{multicols}{2}
The corresponding normal system of equations in the standard notation (\ref{sys_norm}) has the following form:

\begin{equation} \label{Eq__28_}
x'=P;\, \, \, \, \, Z'=Q;\, \, \, \, \, y'=p;\, \, \, \, \, z'=q.
\end{equation}
The necessary condition of the solution's reality \eqref{Eq__26_} can be re-written in the following form:
\begin{equation} \label{Eq__29_}
\bigl(Z^{2} +ex^{2} -\frac{\alpha _{m} }{2} x^{4} \bigr)-\bigl(z^{2} -\varepsilon \mu ^{2} y^{2} +\frac{\beta _{m} }{2} y^{4} \bigr)+\lambda _{m} \ge 0.
\end{equation}
%
%%%%%%%%%%%%%%%%%%%%%%%%%%%%%%%%%%%%%%%%%%%%%%%%%%%%%%%%%%%%%%%%%%%%%%%%%%%%%%%%%%%%%%%%%%%%%%%%%%%%%
%%%%%%%%%%%%%%%%%%%%%%%%%%%%%%%%%%%%%%%%%%%%%%%%%%%%%%%%%%%%%%%%%%%%%%%%%%%%%%%%%%%%%%%%%%%%%%%%%%%%%

\subsection{Areas of the Reality of the Solution and the Motion in the Neighbourhood of Energy Hypersurfaces}
As was mentioned above, the unique property of the considered system is change of the topology of the phase space as a consequence of appearance of the ranges in it, where the motion is not possible. These ranges stand out by the condition of non-negativeness of the effective total energy \eqref{E_m} while ranges that are available for the phase trajectories are defined by the condition (\ref{Eq__26_}). The null effective energy hypersurfaces $S_{3}^{0} \subset {\mathbb R}_{4} $, splitting the phase space to the admitted and forbidden regions of the dynamic variables, are described by the equations:
\begin{eqnarray} \label{Eqs12_}
\mathcal{E}_{m}(\Phi,Z,\varphi,z) =0\Rightarrow \nonumber\\
\frac{Z^{2} }{2} -V(\Phi )-\left(\frac{z^{2} }{2} -v(\varphi )\right)+\Lambda _{m} =0.
\end{eqnarray}
Let us notice that as a consequence of definition \eqref{Eqs6_} renormalized value of the cosmological constant $\Lambda _{m} $ can take generally speaking negative values as well. Let us notice then that the field equations \eqref{Eq__17_} -- \eqref{Eq__18_} have the form of equations of free oscillations in a field of potential of the 4th order
\begin{equation} \label{Eqs13_}
\ddot{x}+k\dot{x}+\frac{\partial V}{\partial x} =0
\end{equation}
with a nonlinear ``coefficient of friction'':
\[k=\sqrt{3\mathcal{E}_{m}(\Phi,Z,\varphi,z)} \equiv \sqrt{3(\mathcal{E}_{c} (\Phi ,Z)+\mathcal{E}_f (\varphi ,z)+\Lambda _{m} )} .\]
According to the theory of oscillations the corresponding dynamic system, losing its total energy as a consequence of dissipative process corresponding to the friction force in \eqref{Eqs13_}, should go down with time to the minimum of the potential energy if it exists; it should be rolling ``down'' infinitely in the case if does not. The specific character of the problem is concluded in, first of all, the significant dependence of the friction force on the total energy of the system, and second, in the nonlinear coupling of the subsystems by the ``friction force'' and third, in the factor of the negativeness of the kinetic energy of the phantom component.
Let us consider the motion of the null effective energy on the hypersurface \eqref{Eqs12_}. The equations \eqref{Dyn_sys} on this hypersurface take the following form:
\begin{equation} \label{Eqs15_}
\begin{array}{l} {\Phi '=Z;} \\ {Z'=-e\Phi +\alpha _{m} \Phi ^{3} ;} \\ {\varphi '=z;} \\ {z'=\varepsilon \mu ^{2} \varphi -\beta _{m} \varphi ^{3} ,} \end{array}
\end{equation}
The right parts of the even equations \eqref{Eqs15_} are derivatives of the corresponding potential functions with respect to scalar fields. Actually, differentiating the relations \eqref{V}--\eqref{v}, we find:
\[\frac{dV}{d\Phi } =e\Phi -\alpha _{m} \Phi ^{3} ;{\rm \; \; \; \; }\frac{dv}{d\varphi } =\varepsilon \mu ^{2} \varphi -\beta _{m} \varphi ^{3}.\]
Thus, multiplying even equations \eqref{Eqs15_} by Z and z we correspondingly obtain the integrals of the total energy at motion along the hypersurface $S_{3}^{0} $:
\begin{eqnarray} \label{Eqs16_}
\frac{Z^{2} }{2} -V(\Phi )\equiv\mathcal{E}_{c}=\mathrm{Cons}t\nonumber\\
\frac{Z^{2} }{2} -\frac{\alpha _{m} }{4} \left(\Phi ^{2} -\frac{e}{\alpha _{m} } \right)^{2} ; \nonumber\\
\frac{z^{2} }{2} -v(\varphi )\equiv -\mathcal{E}_f =\mathrm{Const}\\
=\frac{z^{2} }{2} +\frac{\beta _{m} }{4} \left(\varphi ^{2} -\frac{\varepsilon \mu }{\beta _{m} } \right)^{2}.\nonumber
\end{eqnarray}
This fact exactly shows that we deal with free oscillations. As a consequence of \eqref{Eqs12_} the integrals of the total energy on this trajectory should be coupled by the following relation:
\begin{equation} \label{Eqs17_}
\mathcal{E}_{c} +\mathcal{E}_f +\Lambda _{m} =0.
\end{equation}
Thus we can state that the phase trajectories on the null energy surface which are described by the equations \eqref{Eqs15_} with the integrals of the total energy \eqref{Eqs16_} -- \eqref{Eqs17_} are the exact solution of the complete equations of motion \eqref{Dyn_sys}. To obtain a certain phase trajectory one of the constants ($\mathcal{E}_{c} $ or $\mathcal{E}_f $) should be set while the second one should be obtained from the relation \eqref{Eqs17_}. Thus, the desired phase trajectories are cross-sections of the surface of null effective energy \eqref{Eqs12_}. We can also obtain explicit solutions of the field equations on the surface \eqref{Eqs12_}, integrating \eqref{Eqs16_}:
\begin{eqnarray}
{\Phi =\Phi _{0} \pm \int\limits_{0}^{\tau }\sqrt{2E_{c} +\frac{\alpha _{m} }{2} \left(\Phi ^{2} -\frac{e}{\alpha _{m} } \right)^{2} }  ;} \nonumber\\
 {\varphi =\varphi _{0} \pm \int\limits_{0}^{\tau }\sqrt{2E_f -\frac{\beta _{m} }{2} \left(\varphi ^{2} -\frac{\varepsilon \mu }{\beta _{m} } \right)^{2} }  .} \nonumber
\end{eqnarray}
The result of integration is expressed by means of elliptical functions. In the case of residence of the attractive centers inside the forbidden regions it is expected that the phase trajectories will asymptotically adhere to the surfaces of null effective energy while when saddle points can be found inside these regions it is expected that the phase trajectories will repulse from the surface of null effective energy.

Let us notice, \label{ZamGamma}however, another circumstance which qualitatively differs the cosmological model with asymmetric scalar doublet from the corresponding model with a single scalar field, considered above. The hypersurface of the phase space $S^0_3\in \mathbb{R}_4$ \eqref{Eqs12_} is defined through the parameters of the field model of the scalar doublet $\{e,\varepsilon,\alpha_m,\beta_m,\mu,\lambda_m\}$ and does not depend on the time variable $\tau$.
However, the intersections of the 2-dimensional phase planes of the single fields $\Sigma_\Phi=\{\Phi,Z\}$  and $\Sigma_\varphi=\{\varphi,z\}$ with hypersurface $S^0_3$ can be 2-dimensional curves $\Gamma_\Phi$ and $\Gamma_\varphi$ (closed or open ones), essentially depending on the values of the dynamic variables of another scalar field and therefore depending on the time variable:
\begin{eqnarray}\label{Gamma}
\Gamma_\Phi(\tau): \; \mathcal{E}_{m}(\Phi,Z,\tau)=0\Rightarrow\mathcal{E}_{m}(\Phi,Z,\varphi(\tau),z(\tau)) =0;\nonumber\\
\Gamma_\varphi(\tau): \;\mathcal{E}_{m}(\varphi,z,\tau)=0\Rightarrow\mathcal{E}_{m}(\Phi(\tau),Z(\tau),\varphi,z) =0.
\end{eqnarray}
As a consequence of \eqref{Gamma} the topology of 2-dimensional phase subsrufaces $\Sigma_\Phi$ and $\Sigma_\varphi$ can significantly change with time looking through all the cases considered in the previous article \cite{Part1}. This factor is essentially new and significant for the cosmological model.
\subsection{The Singular Points of the Dynamic System\label{ST}}
\noindent The singular points of the dynamic system are defined by the system of algebraic equations (see e.g. \cite{Bautin,Bogoyav}):
\begin{equation} \label{Eq__30_}
M:\quad F_{i} (x_{1} ,\ldots ,x_{n} )=0,\quad i=\overline{1,n}.
\end{equation}
According to \eqref{Eq__27_} and \eqref{Eq__30_} these points are defined by the following system of equations:
\begin{equation} \label{Eq__31_}
\begin{array}{l} {Z=0,\quad z=0;} \\ {x(e-\alpha _{m} x^{2} )=0;} \\ {y(\varepsilon \mu ^{2} -\beta _{m} y^{2} )=0.} \end{array}
\end{equation}
\underline{Note 1.} \label{Zam1}Let us notice the following important circumstance. In the case when in certain singular point $M_i$ the reality condition is fulfilled \eqref{Eq__29_}, then, basically the phase trajectories of the dynamic system \eqref{sys_norm} -- \eqref{Eq__27_} can come into such singular point or come out from it. In case singular point is situated in the forbidden range of the phase space, the phase trajectories of the dynamic system can't pass through that singular point and only can be attracted to the boundary of the forbidden range or repulse from it depending on the character of the singular point.\\
\par Thus, as was noted above, the dynamic system \eqref{sys_norm} -- \eqref{Eq__27_} has the following 9 singular points.
\\
\noindent 1. \textit{$M_{0} $}: The system of algebraic equations \eqref{Eq__30_} always has the following trivial solution at any values of $\alpha _{m} $ and $\beta _{m} $ :
\begin{equation} \label{Eq__32_}
x=0;\quad Z=0;\quad y=0;\quad z=0\Rightarrow M_{0} :(0,0,0,0).
\end{equation}
Substituting the obtained solution \eqref{Eq__32_} into the condition \eqref{Eq__29_}, we find the necessary condition of reality of the solutions in a singular point:
\[\lambda _{m} \ge 0.\]
\noindent 2. \textbf{$M_{01} $}, \textbf{$M_{02} $}: We have two more solutions symmetrical in $\varphi $ at any values of $\alpha _{m} $ and $\varepsilon \beta _{m} >0$:
\begin{eqnarray} \label{Eq__34_}
x=0;\quad Z=0;\quad y_{\pm } =\pm \frac{\mu }{\sqrt{\varepsilon \beta _{m} } }\quad (\varepsilon \beta _{m} >0) ;\nonumber\\
z=0\quad \Rightarrow M_{01} (0,0,|y_{\pm } |,0);\quad M_{02} (0,0,-|y_{\pm } |,0).
\end{eqnarray}
The necessary condition of reality of the solutions in singular points \textbf{$M_{01} $}, \textbf{$M_{02} $} is:
\begin{equation}\label{sigma1^2}
\sigma_1^2\equiv \frac{3}{4}\biggl(\lambda _{m}+\frac{\mu ^{4} }{2\beta _{m} }\biggr) + \ge 0.
\end{equation}
\noindent 3.~\textbf{$M_{10} $}, $M_{20} $: We have two more solutions which are symmetrical in $\Phi $ at any $\beta _{m} $ and $e\alpha _{m} >0$:
\begin{eqnarray} \label{Eq__36_}
x_{\pm } =\pm \frac{1}{\sqrt{e\alpha _{m} } } ;\quad Z=0;\quad y=0;\quad z=0\quad (e\alpha_m>0) \nonumber\\
\Rightarrow M_{10} (|x_{\pm } |,0,0,0);\quad M_{20} (-|x_{\pm } |,0,0,0).
\end{eqnarray}
The necessary condition of reality of the solutions in singular points  \textbf{$M_{10} $}, $M_{20} $ is:
\begin{equation}\label{sigma2^2}
\sigma_2^2\equiv\frac{3}{4}\biggl(\frac{1}{2\alpha _{m} } +\lambda _{m}\biggr) \ge 0.
\end{equation}
\noindent 4.~$M_{11} $, $M_{12} $, $M_{21} $, $M_{22} $: We have 4 more solutions symmetrical in $\Phi $ and $\varphi $ at $e\alpha _{m} >0$ and $\varepsilon \beta _{m} >0$ :
\begin{equation} \label{Eq__38_}
\begin{array}{l}
\displaystyle{x_{\pm } =\pm \frac{1}{\sqrt{e\alpha _{m} } } ;\quad Z=0;\quad y_{\pm } =\pm \frac{\mu }{\sqrt{\varepsilon \beta _{m} } } ;\quad z=0\quad \Rightarrow } \\[12pt]
{M_{11} (|x_{\pm } |,0,|y_{\pm } |,0);{\rm }M_{12} (|x_{\pm } |,0,-|y_{\pm } |,0);} \\[12pt]
{M_{21} (-|x_{\pm } |,0,|y_{\pm } |,0);M_{22} (-|x_{\pm } |,0,-|y_{\pm } |,0).} \end{array}
\end{equation}
The necessary condition of reality of the solutions in singular points $M_{11} $, $M_{12} $, $M_{21} $, $M_{22} $ is:
\begin{equation}\label{sigma3^2}
\sigma_3^2\equiv\frac{3}{4}\biggl(\frac{1}{2\alpha _{m} } +\frac{\mu ^{4} }{2\beta _{m} } +\lambda _{m}\biggr) \ge 0.
\end{equation}
\subsection{The Character of Singular Points of the Dynamic System of Assymetrical Scalar Doublet}
A minimal character of interaction of the doublet's components univalently lead to block-diagonal structure of the \eqref{Eq__27_} dynamic system's matrix \footnote{see \cite{Bautin,Bogoyav}}, which has the following form at $Z=z=0$ (\ref{Eq__31_}):
\[
\hskip 0mm A_M\equiv\left\| \frac{\partial F_{i} }{\partial x_{k} } \right\|_M =\left(\begin{array}{cccc} {0} & {1} & {0} & {0} \\ {\displaystyle\frac{\partial Q}{\partial x} } & {\displaystyle\frac{\partial Q}{\partial Z} } & {0} & {0} \\ {0} & {0} & {0} & {1} \\ {0} & {0} & {\displaystyle\frac{\partial q}{\partial y} } & {\displaystyle\frac{\partial q}{\partial z} } \end{array}\right)_M\!\!\!.
\]
The determinant of this matrix is equal to:
\[
\Delta (A)=\frac{\partial Q}{\partial x} \frac{\partial q}{\partial y} .
\]
\hrule\vskip 4pt
\noindent\textbf{Notice 2.}
\label{Zam2}Let us notice that since all the dynamic variables are real values, all functions $Q,q$ with their partial derivatives in the admitted regions of the phase space are also real values. However, in the forbidden ranges of the phase space, i.e. in ranges with negative effective energy $\mathcal{E}_m<0$, the derivatives over the dynamic variables can become imaginary values. This means that this singular point is situated in the inaccessible region of the phase space. Let us notice that condition of reality in the matrix of the dynamic system can be even simultaneously violated in the derivatives of type $\partial Q/\partial Z, \partial Q/\partial z, \partial q/\partial Z, \partial q/\partial z $.\vskip 4pt
\hrule\vskip 12pt

Let us consider equations for eigenvectors $\mathbf{u}_i$ and eigenvalues $k_i$ of the dynamic system's matrix:
\begin{eqnarray}\label{eigen_vectors}
\bigl(A_M-k_iE\bigr)\mathbf{u}_i=0;\\
\label{eigen_values}
\mathrm{Det}\bigl(A_M-k_iE\bigr)=0,
\end{eqnarray}
where $E$ is an identity matrix. Due to block-diagonal structure of the dynamic system's matrix $A_M$ its eigenvalues are defined by the characteristic equations in corresponding planes while eigenvectors $\mathbf{u}^{(M)}_k$, corresponding to these eigenvalues, lie pairwise in different phase planes: $\{\mathbf{u}^{(0)}_1,\mathbf{u}^{(0)}_1\}\in \Sigma_\Phi$,
$\{\mathbf{u}^{(M)}_3,\mathbf{u}^{(M)}_4\}\in \Sigma_\varphi$. This fact allows to significantly simplify the qualitative analysis of the phase trajectories in the neighbourhood of the singular point $M^\alpha$ and reduce it to enumeration of combinations of the dynamic system's oscillations in 2-dimensional planes $\Sigma_\Phi, \Sigma_\varphi$. \label{Zam2}If a singular point is situated in the admitted region, then as a consequence of reality of the elements of the dynamic system's matrix, each of its complex eigenvalues $k$ should be corresponded by a complex conjugated value $\overline{k}$, so that $k\overline{k}=|k|^2>0$. If a singular point is situated in the inaccessible region of the phase space, the last condition may not hold. In this case the conclusions of the qualitative theory are only conditionally applicable to the extent that the prohibited region's radius is small. A certain behavior of the phase trajectory in these cases should be specified with the help of numerical integration of the dynamic equations. According to the qualitative theory of differential equations (see \cite{Bautin,Bogoyav}) the radius-vector of the phase trajectory $\mathrm{r}(\tau)=(x_1(\tau),\ldots,x_n(\tau))$ in the neighbourhood of the singular point $M^\alpha(x^{(\alpha_1)},\ldots,x^{(\alpha)}_n)$ is decribed by the following equation:
\begin{equation}\label{r(t)}
\mathbf{r}(t)\simeq \mathbf{r}^{(\alpha)}+\Re\biggl(\sum\limits_{j=1}^n C_j \mathbf{u}^{(\alpha)}_j e^{ik^{(\alpha)}_j\tau}\biggr),
\end{equation}
where $C_j$ are arbitrary constants which are defined by the initial conditions, $k^{(\alpha)}_j$ are eigenvalues of the dynamic system's matrix $A(M_\alpha)$, $\mathbf{u}^{(\alpha)}_j$ are eigenvectors of this matrix corresponding to eigenvalues $k^{(\alpha)}_j$. In the cases when a singular point is inaccessible, the evaluation formula (\ref{r(t)}) is nevertheless good enough approximation of the phase trajectory. We will rely on this evaluation in the cases when the results of the standard qualitative theory which could be suitable for real matrices of the dynamic system, are absent.

Let us briefly state the results of the qualitative analysis of the dynamic system \eqref{Dyn_sys}. First, these calculations show that all the singular points of the dynamic system are split into 4 groups where the character of points inside each group is the same.

\subsubsection{The Characteristic Equation and the Qualitative Analysis In The Neighbourhood of Null Singular Point $M_{0} $ }
the following single singular point is included in the first group:
\[M_0(0,0,0,0); \quad \alpha\in \mathbb{R},\quad \beta\in \mathbb{R}\]
The system's matrix \eqref{Eq__27_} in null singular point \eqref{Eq__32_} at any $\alpha _{m} $ and $\beta _{m}$ takes the following form:
\[
A_{0} \equiv A(M_{0} )=\left(\begin{array}{cccc} {0} & {1} & {0} & {0} \\ {-e} & {-\sqrt{3\lambda _{m} } } & {0} & {0} \\ {0} & {0} & {0} & {1} \\ {0} & {0} & {\varepsilon \mu ^{2} } & {-\sqrt{3\lambda _{m} } } \end{array}\right),
\]
and its determinant is equal to:
\[
\Delta (A_{0} )=-e\varepsilon \mu ^{2} .
\]
As a consequence of the notice made above on page \pageref{Zam2}, point $M_0$ is accessible at $\lambda_m\geq 0$ and is inaccessible at $\lambda_m\leq 0$.
The characteristic equation for matrix $A_{0} $ has the following form:
\[(k^{2} +k\sqrt{3\lambda _{m} } +e)(k^{2} +k\sqrt{3\lambda _{m} } -\varepsilon \mu ^{2} )=0,\]
thus the eigenvalues of the matrix are equal to:
\begin{eqnarray} \label{Eq__44_}
k_{1,2} (M_{0} )=&\displaystyle- \sqrt{\frac{3\lambda _{m}}{4} } \pm\sqrt{\frac{3\lambda _{m}}{4} -e} ;\nonumber \\[8pt]
k_{3,4} (M_{0} )=&\displaystyle -\sqrt{\frac{3\lambda _{m}}{4} } \pm\sqrt{\frac{3\lambda _{m}}{4}  +\varepsilon \mu ^{2} } ,
\end{eqnarray}
herewith it is
\begin{equation} \label{Eq__45_}
\begin{array}{l} k_{1} (M_{0} )\cdot k_{2} (M_{0} )=e;\; k_{3} (M_{0} )\cdot k_{4} (M_{0} )=-\varepsilon \mu ^{2}; \\
{k_{1} (M_{0} )\cdot k_{2} (M_{0} )\cdot k_{3} (M_{0} )\cdot k_{4} (M_{0} )=\Delta (A_{0} ).} \end{array}
\end{equation}
Thus, according to the qualitative theory of differential equations, singular point $M_0$ can have the following character depending on the parameters of the field model (Table \ref{tab1}, \ref{tab2}).

\begin{center}
\tabcaption{ \label{tab1}  The character of the singular point $M_0$ in the plane $\Sigma_\Phi$.}
\footnotesize
\begin{tabular*}{80mm}{c@{\extracolsep{\fill}}ccc}
\toprule $\lambda_m$ & $e$  & $k$  & \ type \\
\hline
$0$ & $+1$ & $\pm i$ & Center \\[8pt]
 & $-1$   & $ \pm 1 $ & Saddle \\
 \hline
$\lambda_m>0$ & $-1$ & $k_1>0;k_2<0$ & Saddle\\[8pt]
$\lambda_m>4/3$ & $+1$ & $k_1<0;k_2<0$ & \parbox{1.8cm}{Attractive Node} \\[8pt]
$0<\lambda_m<4/3$ & $+1$ & $\Re(k)<0$ & \parbox{1.8cm}{Attractive Focus} \\[8pt]
\hline
$\lambda_m<0 $ & $+1 $  & $\Re(k_1)=\Re(k_2)=0 $ & \parbox{1.8cm}{Inaccessible Center$^{*)}$} \\
$-4/3<\lambda_m<0 $ & $-1 $  & $\Re(k_1)=-\Re(k_2) $ & \parbox{1.8cm}{Inaccessible Saddle$^{**)}$} \\
$\lambda_m<-4/3 $ & $-1 $  & $\Re(k_1)=\Re(k_2)=0 $ & \parbox{1.8cm}{Inaccessible Center$^{*)}$} \\
\bottomrule
\end{tabular*}
\end{center}
$^{*)}$ \emph{Note}. Let us consider examples of such trajectories, selecting corresponding parameters in (\ref{r(t)}).
\begin{flushleft}
\includegraphics[width=8.5cm,height=4cm]{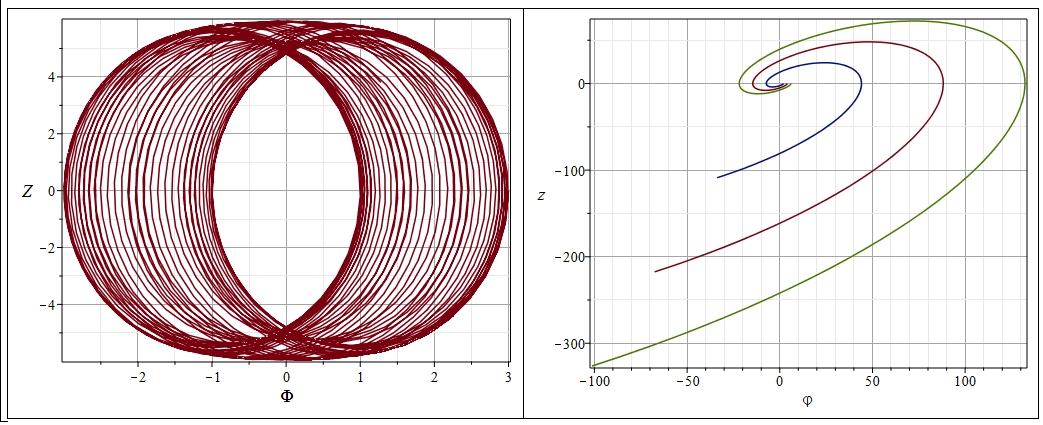}
\figcaption{\label{ris2} Examples of phase trajectories  (\ref{r(t)}) in the neighbourhood of singular points. On the left-hand side -- inaccessible center$^{*)}$ in the plane $\Sigma_\Phi$: $\lambda_m=-1,e=1$; on the right-hand side -- inaccessible saddle$^{**)}$ in the plane $\Sigma_\varphi$: (\ref{r(t)}) $\lambda_m=-1,\varepsilon=1, \mu=1$.}
\end{flushleft}
%
%c@{\extracolsep{\fill}}
%
\begin{center}
\tabcaption{ \label{tab2} The character of the singular point $M_0$ in the plane $\Sigma_\varphi$.}
\footnotesize
\begin{tabular*}{80mm}{llll}
\toprule $\lambda_m$ & $\varepsilon$  & $k$  & \ type \\
\hline
$0$ & $+1$ & $\pm \mu$ & Saddle \\
 & $-1$   & $ \pm i\mu $ & Center \\[6pt]
\hline\\[-4pt]
 $\lambda_m>0$ & $+1$ & $k_3>0;k_4<0$ & Saddle\\
 $\lambda_m>4/3\mu^2$ & $-1$ & $k_3<0;k_4<0$ & \parbox{1.8cm}{Attrac\-tive node} \\[0pt]
$0<\lambda_m<4/3\mu^2$ & $-1$ & $\Re(k)<0$ & \parbox{1.8cm}{Attrac\-tive focus} \\[6pt]
\hline\\[-2pt]
$-4/3\mu^2<\lambda_m<0$ &  $+1$ & $\Re(k_4)=-\Re(k_3)$ & \parbox{1.8cm}{Inaccessible saddle$^{**)}$}\\
$\lambda_m<-4/3\mu^2$ &  $+1$ &  $\Re(k_3)=\Re(k_4)=0 $ & \parbox{1.8cm}{Inaccessible center$^{*)}$} \\
$\lambda_m<0$ &  $-1$ &  $\Re(k_3)=\Re(k_4)=0 $ & \parbox{1.8cm}{Inaccessible center$^{*)}$} \\
\bottomrule
\end{tabular*}
\vskip 12pt
\end{center}
%
%ris2

\subsubsection{The Characteristic Equation and the Qualitative Analysis in the Neighbourhood of Singular Points $M_{01}, M_{02}$ }
Two symmetrical singular points $M_{01}, M_{02}$ (\ref{Eq__34_}) are included in the second group providing $\varepsilon\beta_m>0$ and $\forall\alpha_m$ :
\[M_{01}\biggl(0,0,\frac{\mu}{\sqrt{\varepsilon\beta_m}},0\biggr); \quad M_{02}\biggl(0,0,-\frac{\mu}{\sqrt{\varepsilon\beta_m}},0\biggr),\]
the dynamic system's matrices $A_{01}$ and $A_{02}$ in these points coincide and are equal to:
\[
A_{01} \equiv A(M_{01} )=A(M_{02})=
\left(
\begin{array}{cccc}
{0} & {1} & {0} & {0} \\
 -e & 2\sigma_1 & {0} & {0} \\
 {0} & {0} & {0} & {1} \\
 {0} & {0} & -2\varepsilon \mu ^{2} & 2\sigma_1
 \end{array}
 \right),
\]
where $\sigma_1^2$ is defined by formula (\ref{sigma1^2}). As a consequence of the note made on page \pageref{Zam2} the following condition on the model's parameters should be fulfilled  \emph{in the accessible singular points}:
\begin{equation}
\sigma_1^2\geq 0\Rightarrow
\left\{\begin{array}{ll}
\beta_m>0, & \lambda_m=0;\\
\displaystyle\beta_m\in\biggl(-\infty,-\frac{\mu^4}{2\lambda_m}\biggr)\cup (0,+\infty), & \lambda_m>0.\\
\end{array}\right.
\end{equation}
In the opposite case the singular points $M_{01}, M_{02}$ are situated in inaccessible regions. The eigenvalues of the matrix  $A_{01}$ are equal to:
\begin{eqnarray} \label{Eq__44_}
k_{1,2} (M_{01} )=&\displaystyle -\sigma_1\pm\sqrt{\sigma_1^2 -e} ;\nonumber \\[8pt]
k_{3,4} (M_{01} )=&\displaystyle -\sigma_1 \pm\sqrt{\sigma_1^2- 2\varepsilon\mu^2},
\end{eqnarray}
where it is
\begin{equation}\label{Delta01}
\Delta (A_{01})=2e\varepsilon \mu ^{2};\; k_1k_2=e;\; k_3k_4=2\varepsilon\mu^2.
\end{equation}
%
%%%%%%%%%%%%%%%%%%%%%%%%%%%%%%%%%%%%%%%%%%%%%%%%%%%%%%%%%%%%%%%%%%%%%%%%%%%%%%%
%
Apparently, the role of parameter $3\lambda_m/4$ in the case considered above is played by the combined parameter $\sigma_1^2$. As a result, the following types of these points are possible (Table \ref{tab3}, \ref{tab4} )
\begin{center}
\tabcaption{ \label{tab3}  The character of the singular points $M_{01},M_{02}$ in the plane $\Sigma_\Phi$.}
\footnotesize
\begin{tabular*}{80mm}{c@{\extracolsep{\fill}}ccc}
\toprule $\sigma_1^2$ & $e$  & $k$  & \ type \\
\hline
$0$ & $+1$ & $\pm i$ & Center \\
 & $-1$   & $ \pm 1 $ & Saddle \\
 \hline
$\sigma_1^2>0$ & $-1$ & $k_1>0;k_2<0$ & Saddle\\
$\sigma_1^2>1$ & $+1$ & $k_1<0;k_2<0$ & \parbox{2cm}{Attractive node} \\
$0<\sigma_1^2<1$ & $+1$ & $\Re(k)<0$ & \parbox{2cm}{Attractive focus} \\[6pt]
\hline\\[-2pt]
$\sigma_1^2<0 $ & $+1$ & $\Re(k)=0$ & \parbox{2cm}{Inaccessible center$^{*)}$}\\
$-1<\sigma_1^2<0 $ & $-1$ & $\mathrm{R}(k_1)=-\mathrm{R}(k_2)$ & \parbox{2cm}{Inaccessible saddle$^{**)}$}\\
$\sigma_1^2<-1 $ & $-1$ & $\Re(k)=0$ & \parbox{2cm}{Inaccessible center$^{*)}$}\\
\bottomrule
\end{tabular*}
\end{center}
\vspace{12pt}
\begin{center}
\tabcaption{ \label{tab4} The character of the singular points $M_{01},M_{02}$ in the plane $\Sigma_\varphi$.}
\footnotesize
\begin{tabular*}{80mm}{c@{\extracolsep{\fill}}ccc}
\toprule $\sigma_1^2$ & $\varepsilon$  & $k$  & \ type \\
\hline
$0$ & $+1$ & $\pm i\sqrt{2}\mu$ & Center \\
 & $-1$   & $ \pm \sqrt{2}\mu $ & Saddle \\
 \hline
 $\sigma_1^2>0$ & $-1$ & $k_3>0;k_4<0$ & Saddle\\
 $\sigma_1^2>2\mu^2$ & $+1$ & $k_3<0;k_4<0$ & \parbox{2cm}{Attractive node} \\
$0<\sigma_1^2<2\mu^2$ & $+1$ & $\Re(k)<0$ & \parbox{2cm}{Attractive focus} \\
\hline
$\sigma_1^2<0$ &  $+1$ & $\Re(k)=0$ & \parbox{2cm}{Inaccessible center$^{*)}$}\\
$\sigma_1^2<-2\mu^2$ &  $-1$ & $\Re(k)=0$ & \parbox{2cm}{Inaccessible center$^{*)}$}\\
$-2\mu^2<\sigma_1^2<0$ \; &  $-1$ & $\mathrm{R}(k_3)=-\mathrm{R}(k_4)$ & \parbox{2cm}{Inaccessible saddle$^{**)}$}\\
\bottomrule
\end{tabular*}
\end{center}
\subsubsection{Singular Points $M_{10}$ and $M_{20}$:}
The third group of singular points comprises points (\ref{Eq__36_})  $M_{10}$ and $M_{20}$ provided $e\alpha_m>0$
\[
M_{10}  \biggl(\frac{1}{\sqrt{e\alpha _{m}}},0,0,0\biggr);\quad M_{20}\biggl(-\frac{1}{\sqrt{e\alpha _{m}}},0,0,0\biggr).
\]
The matrices of the dynamic system  $A_{01}$ and $A_{02}$ in these points coincide and are equal to:
\[
A_{10} \equiv A(M_{10} )=A(M_{20})=
\left(
\begin{array}{cccc}
{0} & {1} & {0} & {0} \\
 2e & -2\sigma_2 & {0} & {0} \\
 {0} & {0} & {0} & {1} \\
 {0} & {0} & \varepsilon \mu ^{2} & -2\sigma_2
 \end{array}
 \right),
\]
where $\sigma_2^2$ is defined by formula (\ref{sigma2^2}). The following condition on the parameters of the model should be fulfilled \emph{in the accessible singular points}:
\begin{equation}
\sigma_1^2\geq 0\Rightarrow
\left\{\begin{array}{ll}
\alpha_m>0, & \lambda_m=0;\\
\displaystyle\alpha_m\in\biggl(-\infty,-\frac{1}{2\lambda_m}\biggr)\cup (0,+\infty), & \lambda_m>0.\\
\end{array}\right.
\end{equation}
In the opposite case, singular points $M_{10}, M_{20}$ are situated in inaccessible regions. The eigenvalues of the matrix $A_{10}$ are equal to:
\begin{eqnarray} \label{Eq__44a_}
k_{1,2} (M_{10} )=&\displaystyle -\sigma_2\pm\sqrt{\sigma_2^2 +2e} ;\nonumber \\[8pt]
k_{3,4} (M_{10} )=&\displaystyle -\sigma_2 \pm\sqrt{\sigma_2^2+\varepsilon\mu^2},
\end{eqnarray}
where it is
\begin{equation}\label{Delta10}
\Delta (A_{10})=2e\varepsilon \mu ^{2};\; k_1k_2=e;\; k_3k_4=2\varepsilon\mu^2.
\end{equation}
%
%%%%%%%%%%%%%%%%%%%%%%%%%%%%%%%%%%%%%%%%%%%%%%%%%%%%%%%%%%%%%%%%%%%%%%%%%%%%%%%
%
Apparently, the role of the parameter $\sigma_1$ in the considered above case is played by the parameter $\sigma_2$. As a result, the following types of these points are possible (Table \ref{tab5}, \ref{tab6})

\begin{center}
\tabcaption{ \label{tab5}  The character of the singular points $M_{10},M_{20}$ in the plane $\Sigma_\Phi$.}
\footnotesize
\begin{tabular*}{80mm}{c@{\extracolsep{\fill}}ccc}
\toprule $\sigma_2^2$ & $e$  & $k$  & \ type \\
\hline
$0$ & $+1$ & $\pm \sqrt{2}$ & Saddle \\
 & $-1$   & $ \pm i\sqrt{2} $ & Center \\
 \hline
$\sigma_2^2>0$ & $+1$ & $k_1>0;k_2<0$ & Saddle\\
$\sigma_2^2>2$ & $-1$ & $k_1<0;k_2<0$ & \parbox{2cm}{Attractive node} \\
$0<\sigma_1^2<2$ & $-1$ & $\Re(k)<0$ & \parbox{2cm}{Attractive focus} \\
\hline
$\sigma_2^2<0 $ & $-1$ & $\Re(k)=0$ & \parbox{2cm}{Inaccessible center$^{*)}$}\\
$-2<\sigma_2^2<0\; $ & $+1$ & $\mathrm{R}(k_1)=-\mathrm{R}(k_2)$ & \parbox{2cm}{Inaccessible saddle$^{**)}$}\\
$\sigma_2^2<-2 $ & $+1$ & $\Re(k)=0$ & \parbox{2cm}{Inaccessible center$^{*)}$}\\
\bottomrule
\end{tabular*}
\end{center}
\begin{center}
\tabcaption{ \label{tab6}  The character of the singular points $M_{10},M_{20}$ in the plane $\Sigma_\varphi$.}
\footnotesize
\begin{tabular*}{80mm}{c@{\extracolsep{\fill}}ccc}
\toprule $\sigma_2^2$ & $\varepsilon$  & $k$  & \ type \\
\hline
$0$ & $+1$ & $\pm \mu$ & Saddle \\
 & $-1$   & $ \pm i\mu $ &  Center\\
 \hline
 $\sigma_2^2>0$ & $+1$ & $k_3>0;k_4<0$ & Saddle\\
 $\sigma_2^2>\mu^2$ & $-1$ & $k_3<0;k_4<0$ & \parbox{2cm}{Attractive node} \\
$0<\sigma_2^2<\mu^2$ & $-1$ & $\Re(k)<0$ & \parbox{2cm}{Attractive focus} \\
\hline
$\sigma_2^2<0$ &  $-1$ & $\Re(k)=0$ & \parbox{2cm}{Inaccessible center$^{*)}$}\\
$-\mu^2<\sigma_2^2<0$ &  $+1$ & $\mathrm{R}(k_3)=-\mathrm{R}(k_4)$ & \parbox{2cm}{Inaccessible saddle$^{**)}$}\\
$\sigma_2^2<-\mu^2$ &  $+1$ & $\Re(k)=0$ & \parbox{2cm}{Inaccessible center$^{*)}$}\\
\bottomrule
\end{tabular*}
\end{center}
\subsubsection{Singular Points $M_{11}$,  $M_{22}$,  $M_{12}$ and $M_{21}$}
Finally, the forth group of singular points comprise points (\ref{Eq__38_})  $M_{11}$,  $M_{22}$,  $M_{12}$ and $M_{21}$ provided $e\alpha_m>0$ and $\varepsilon\beta_m>0$:
\begin{eqnarray} \label{Eq__38_}
M_{11} \biggl(\frac{1}{\sqrt{e\alpha _{m} } }  ,0,\frac{\mu }{\sqrt{\varepsilon \beta _{m} }} ,0\biggr);
M_{12}  \biggl(\frac{1}{\sqrt{e\alpha _{m} } }  ,0,-\frac{\mu }{\sqrt{\varepsilon \beta _{m} } } ,0\biggr);\nonumber\\
M_{21} \biggl(-\frac{1}{\sqrt{e\alpha _{m} } }  ,0,\frac{\mu }{\sqrt{\varepsilon \beta _{m} }} ,0\biggr);
M_{22}  \biggl(-\frac{1}{\sqrt{e\alpha _{m} } }  ,0,-\frac{\mu }{\sqrt{\varepsilon \beta _{m} } } ,0\biggr),\nonumber
\end{eqnarray}
The matrices of the dynamic system  $A_{11}$, $A_{12}$, $A_{21}$ and $A_{22}$  in these points coincide and are equal to:
\[
A_{11} \equiv A(M_{11} )=
\left(
\begin{array}{cccc}
{0} & {1} & {0} & {0} \\
 2e & -2\sigma_3 & {0} & {0} \\
 {0} & {0} & {0} & {1} \\
 {0} & {0} & -2\varepsilon \mu ^{2} & -2\sigma_3
 \end{array}
 \right),
\]
where $\sigma_3^2$ is defined by formula (\ref{sigma3^2}). The following condition on the parameters of the model should be fulfilled \emph{in the accessible singular points}:
\begin{equation}\label{sigma3^2>0}
\sigma_3^2\geq 0\Rightarrow
\lambda_m+\frac{1}{2\alpha_m}+\frac{\mu^4}{2\beta_m}\geq 0.
\end{equation}
In the opposite case, singular points $M_{11}, M_{12}, M_{21}, M_{22}$ are situated in inaccessible regions. The eigenvalues of the matrix $A_{11}$ are equal to:

\begin{eqnarray} \label{Eq__44b_}
k_{1,2} (M_{10} )=&\displaystyle -\sigma_3\pm\sqrt{\sigma_3^2 +2e} ;\nonumber \\[8pt]
k_{3,4} (M_{10} )=&\displaystyle -\sigma_3 \pm\sqrt{\sigma_3^2-2\varepsilon\mu^2},
\end{eqnarray}
where it is
\begin{equation}\label{Delta10}
\Delta (A_{10})=-4e\varepsilon \mu ^{2};\; k_1k_2=-2e;\; k_3k_4=2\varepsilon\mu^2.
\end{equation}
%
%%%%%%%%%%%%%%%%%%%%%%%%%%%%%%%%%%%%%%%%%%%%%%%%%%%%%%%%%%%%%%%%%%%%%%%%%%%%%%%
%
As a result, the following types of these points are possible Tab. \ref{tab7}, \ref{tab8} ).

\begin{center}
\tabcaption{ \label{tab7}  The character of the singular points $M_{11},M_{12},M_{211},M_{22}$ in the plane $\Sigma_\Phi$.}
\footnotesize
\begin{tabular*}{80mm}{c@{\extracolsep{\fill}}ccc}
\toprule $\sigma_3^2$ & $e$  & $k$  & \ type \\
\hline
$0$ & $+1$ & $\pm \sqrt{2}$ & Saddle \\
 & $-1$   & $ \pm i\sqrt{2} $ & Center \\
 \hline
$\sigma_3^2>0$ & $+1$ & $k_1>0;k_2<0$ & Saddle\\
$\sigma_3^2>2$ & $-1$ & $k_1<0;k_2<0$ & \parbox{2cm}{Attractive node} \\
$0<\sigma_3^2<2$ & $-1$ & $\Re(k)<0$ & \parbox{2cm}{Attractive focus} \\
\hline
$\sigma_3^2<0 $ & $-1$ & $\Re(k)=0$ & \parbox{2cm}{Inaccessible center$^{*)}$}\\
$-2<\sigma_3^2<0 $ & $+1$ & $\mathrm{R}(k_1)=-\mathrm{R}(k_2)$ & \parbox{2cm}{Inaccessible saddle$^{**)}$}\\
$\sigma_3^2<-2 $ & $+1$ & $\Re(k)=0$ & \parbox{2cm}{Inaccessible center$^{*)}$}\\
\bottomrule
\end{tabular*}
\end{center}
\begin{center}
\tabcaption{ \label{tab8}  The character of the singular point $M_{11},M_{12},M_{211},M_{22}$ in the plane $\Sigma_\varphi$.}
\footnotesize
\begin{tabular*}{80mm}{c@{\extracolsep{\fill}}ccc}
\toprule $\sigma_3^2$ & $\varepsilon$  & $k$  & \ type \\
\hline
$0$ & $+1$ & $\pm i\sqrt{2} \mu$ & Center \\
 & $-1$   & $ \pm \sqrt{2}\mu $ & Saddle \\
 \hline
 $\sigma_3^2>0$ & $-1$ & $k_3>0;k_4<0$ & Saddle\\
 $\sigma_3^2>2\mu^2$ & $+1$ & $k_3<0;k_4<0$ & \parbox{2cm}{Attractive node} \\
$0<\sigma_3^2<2\mu^2$ & $+1$ & $\Re(k)<0$ & \parbox{2cm}{Attractive focus} \\
\hline
$\sigma_2^2<0$ &  $+1$ & $\Re(k)=0$ & \parbox{2cm}{Inaccessible center$^{*)}$}\\
$-2\mu^2<\sigma_2^2<0$ &  $-1$ & $\mathrm{R}(k_3)=-\mathrm{R}(k_4)$ & \parbox{2cm}{Inaccessible saddle$^{**)}$}\\
$\sigma_1^2<-2\mu^2$ &  $-1$ & $\Re(k)=0$ & \parbox{2cm}{Inaccessible center$^{*)}$}\\
\bottomrule
\end{tabular*}
\end{center}
Thus, we enumerated all the characteristics of singular points in 2-dimensional planes $\Sigma_\Phi$ and $\Sigma_\varphi$. To obtain a general characteristic of each singular point, it is required to multiply their characteristics in the corresponding planes. However, it is necessary to validate the corresponding conditions on the parameters for their consilience. As an example, let us consider consilient cases $\sigma_3^2>2, e=-1$ from the table \ref{tab7} and $\sigma_3^2>0,\varepsilon=-1$ from the table \ref{tab8}. A singular point, being an attractive node in the plane $\Sigma_\Phi$ and a saddle in the plane $\Sigma_\varphi$ correspond to them. However, for the existence of such a singular point it is required that $\alpha_m<0,\beta_m<0$. Now we have to check what situation will the conditions $\sigma_3^2>2$and (\ref{sigma3^2>0}) lead to, in order to clarify in which region does the singular point locate.
\section{The Numerical Simulation of the Dynamic System: $\lambda=0$}
Let us provide the results of numerical integration, demonstrating mentioned peculiarities. Let us right away notice that a great number of the parameters of the cosmological model based on the asymmetrical doublet: $\{e,\varepsilon,\alpha_m,\beta_m,\mu,\lambda_m\}$ (6 in total), makes the enumeration of all  possible variations a too cumbersome problem, therefore we will list only certain ones which are most interesting from our point of view. The main part of the numerical simulation's results can be found in the Appendix for the sake of optimization of the material presentation.
The below presented phase trajectories were obtained with the help of the author's package of programs \texttt{DifEqTools} \cite{DifEqTools}, which is an application to the applied package \texttt{Maple}. Further, for the sake of brevity we denote a set of model's parameters by the ordered list of 6 elements  $\mathbf{P}$, and the initial conditions -- by the ordered list of 4 elements $\mathbf{I}$:
\[ \mathbf{P}\equiv[\alpha_m,\beta_m,e,\varepsilon,\mu,\lambda_m];\quad \mathbf{I}\equiv[\Phi(0),Z(0),\varphi(0),z(0)].\]
Let us notice that the initial value of time $\tau_0$ is quite a relative magnitude -- all the dynamic equations of the model as well as the Friedmann metrics are invariant with respect to transfer of the time reference $\tau\rightarrow \tau+\tau_0$.

\subsection{The Case of Accessibility of All Singular Points}
\begin{equation}\label{param0}
\mathrm{P}=[1,1,1,1,1,0].
\end{equation}
\subsubsection{General Properties of the Phase Space}
Singular points in this case have the following coordinates:
 \begin{eqnarray}\label{M1111}
 M_0(0,0,0,0);& M_{10}(1,0,0,0);& M_{01}(-1,0,0,0);\nonumber\\
 M_{10}(0,0,1,0);& M_{20}(0,0,-1,0);& M_{11}(1,0,1,0);\\
 M_{12}(1,0,-1,0);& M_{21}(-1,0,1,0);& M_{01}(-1,0,-1,0),\nonumber
 \end{eqnarray}
and invariant characteristics $\sigma^2_i$ are equal to:
 \[\sigma^2_1=\frac{3}{8};\quad \sigma^2_2=\frac{3}{8};\quad \sigma^2_3=\frac{3}{4}.\]
Since $e\alpha_m=1>0;\varepsilon\beta_m=1>0$, all 9 singular points of the dynamic system are accessible. The character of these points is shown on the Fig. \ref{ris3}. Here and further the centers are shown by the blue color, attractive nodes - by the red color, saddle points - by the white color. Character of the point in the plane $\Sigma_\Phi$ corresponds to the left half of the circle while $\Sigma_\varphi$ corresponds to the right one. The black color of the circle's boundary corresponds to inaccessible point, blue-green color - to the accessible one.
\begin{center}
\includegraphics[width=6cm]{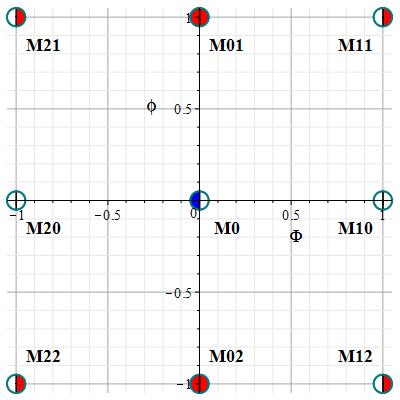}
\figcaption{\label{ris3} The map of singular points in the phase plane $\{\Phi,\varphi\}$ at parameters of the model (\ref{param0}).
}
\end{center}
The Fig. \ref{ris4} illustrates the dependency of the boundaries of the phase space's prohibited ranges projected onto $\Sigma_\Phi$ and $\Sigma_\varphi$ from the value of their dual potentials.
\begin{flushleft}
\includegraphics[width=8.5cm]{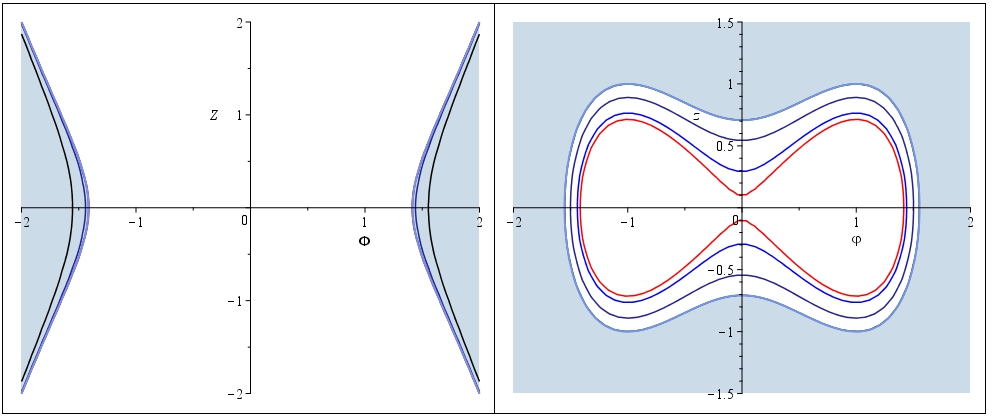}
\figcaption{\label{ris4} The dependency of the prohibited range (highlighted in blue color) in the plane $\Sigma_\Phi$ (on the left) and in the plane $\Sigma_\varphi$ (on the right) on the values of dual potentials at the model's parameters (\ref{param0}). On the left, from inner curves to outer ones it is: $\varphi=0.01;0.1;0.3;1$, the internal ranges are prohibited. On the right, from inner curves to outer ones it is: $\Phi=0.1;0.3;0.6;1$, the external ranges are prohibited.
}
\end{flushleft}
Though the considered case is quite a standard one and does not contain some interesting peculiarities of the model's behavior, let us examine it more in details in order to demonstrate the general properties of the asymmetrical scalar doublet's model.

\subsubsection{The Phase Trajectories of the Dynamic System}
The Fig. \ref{ris5} --\ref{ris8} illustrates the results of numerical simulation of the dynamic system \eqref{Dyn_sys} for the model's parameters \eqref{param0} and initial conditions \eqref{IC0}
\begin{equation}\label{IC0}
\mathbf{I}=[0.7,0.5,0.01,0].
\end{equation}
Here and further green dots correspond to the beginning of the phase trajectory (i.e., the initial values), red ones -- to its final positions, inclined colored crosses denote singular points. It is seen from the Fig. \ref{ris5} that the phase trajectories in the plane $\{\Phi,Z\}$ bounce off the saddle points $M_{10},M_{20}$ and then wind onto the attractive focus $M_0$. Simultaneously with that the phase trajectories in the plane $\{\varphi,z\}$ bounce off the saddle point $M_0$ and wind onto the attractive focuses $M_{01},M_{02}$ (Fig. \ref{ris6}). This situation becomes more clear on the phase diagram in the planes of singular points $\{\Phi,\varphi\}$ (Fig. \ref{ris7}). Thus, the numerical results confirm the outcomes of the qualitative theory represented on the map of singular points (Fig.\ref{ris3}).
\end{multicols}
\ruleup
\begin{center}
\includegraphics[width=18cm,height=6cm]{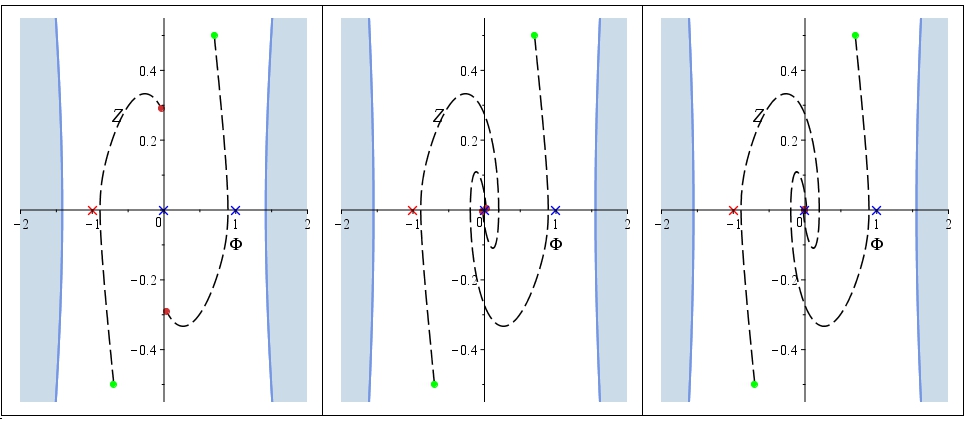}
\figcaption{\label{ris5} The cosmological evolution of the scalar doublet with parameters
 (\ref{param0}) and initial conditions \eqref{IC0} in the ``classical'' plane $\Sigma_\Phi\equiv\{\Phi,Z\}$. The phase diagrams correspond to the time instants (from left to right): $\tau=5;10;20$. The prohibited ranges correspond to the final time instant.
}
\end{center}

\begin{center}
\includegraphics[width=18cm,height=6cm]{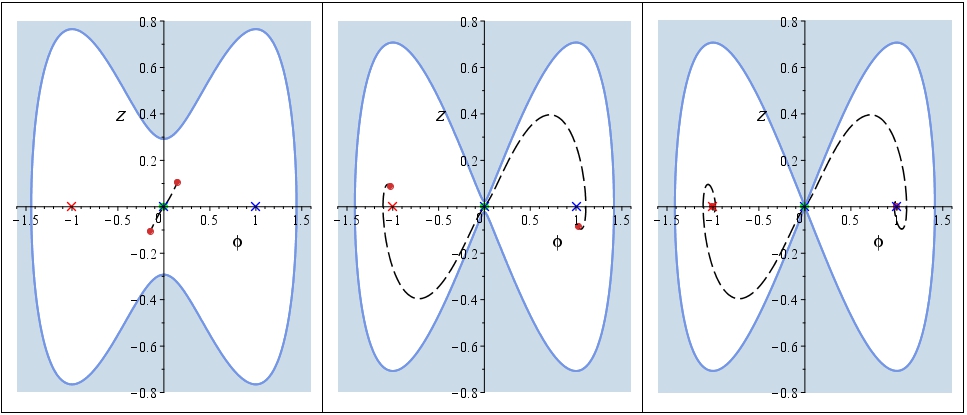}
\figcaption{\label{ris6} The cosmological evolution of the scalar doublet with parameters
 (\ref{param0}) and initial conditions \eqref{IC0} in the phantom plane $\Sigma_\varphi\equiv\{\varphi,z\}$. The phase diagrams correspond to the time instants (from left to right): $\tau=5;10;20$. The prohibited ranges correspond to the final time instant.
}
\end{center}
\begin{center}
\includegraphics[width=18cm,height=6cm]{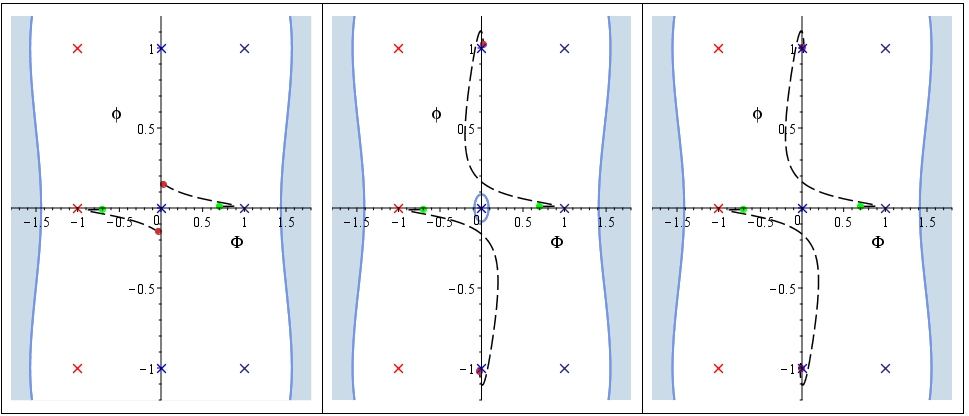}
\figcaption{\label{ris7} The cosmological evolution of the scalar doublet with parameters
 (\ref{param0}) and initial conditions \eqref{IC0} in the plane of potentials $\{\Phi,\varphi\}$. The phase diagrams correspond to the time instants (from left to right): $\tau=5;10;20$.
}
\end{center}

The Fig. \ref{ris8} shows the evolution of the physical characteristics of the cosmological model with parameters \eqref{param0}: dimensionless effective energy $\mathcal{E}_m$ (\ref{E_m}), dimensionless effective pressure $p_m$ (\ref{p_m}) and invariant cosmological acceleration $\Omega$ (see \cite{Part1}):
\begin{eqnarray} \label{Omega}
\Omega= -\frac{1}{2}(1+3\varkappa),\quad
\varkappa\equiv \frac{p_m}{\varepsilon_m} ,
\end{eqnarray}
where  $\varkappa$ is a \emph{barotrope coefficient}. According to generally accepted classification $\varkappa$ takes the value $\varkappa=0$ for non-relativistic matter (in this case it is $\Omega=-1/2$), $\varkappa=1/3$ for the ultrarelativistic matter (in this case $\Omega=-1$), $\varkappa=-1/3$ for the quintessence (in this case it is $\Omega=0$), $\varkappa=-1$ for pure inflation (in this case it is $\Omega=+1$) and $\varkappa<-1$ for dark energy (in this case it is $\Omega>+1$). As is seen from the Fig. \ref{ris8}, all these values in the considered case after the burst come to constant values where
$\varepsilon_m+p_m\to0$ at $\tau\to 0$, which correspond to the inflation on the final stage of the evolution. Thus, the considered case is close enough to the standard scenario with the only difference that late inflation is maintained by the phantom fields rather than by the classical one.

\begin{center}
\includegraphics[width=18cm,height=6cm]{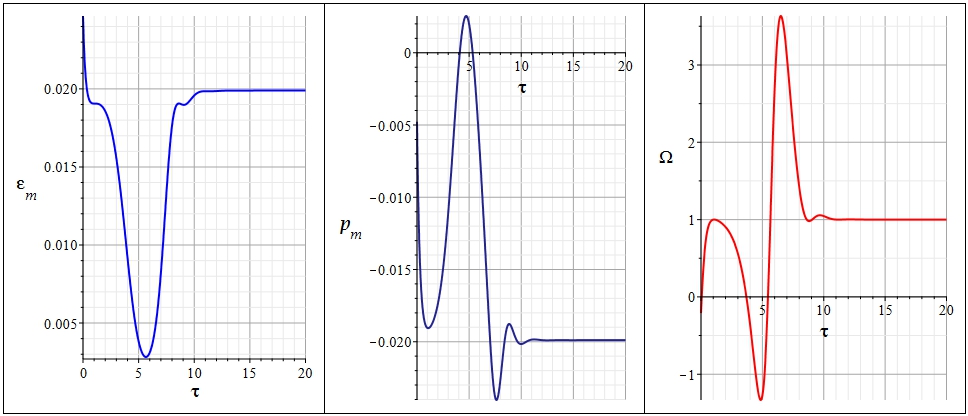}
\figcaption{\label{ris8} The cosmological evolution of the physical characteristics in the cosmological model with parameters
 (\ref{param0}) and initial conditions \eqref{IC0}. From left to right it is: dimensionless effective energy  $\mathcal{E}$ (\ref{E_m}); dimensionless effective pressure $p_m$ (\ref{p_m}); invariant cosmological acceleration $\Omega$ (\ref{Omega}).
 }
\end{center}
%[1,-1,1,-1,1,0],[3,4],1,0.8,[Phi(0)=0,Z(0)=0.7,phi(0)=0.5,z(0)=0]
%
The considered above case corresponded to small values of the phantom potential $\varphi(0)=\pm 0.01$. The increase of the initial value of the phantom potential can significantly change the cosmological scenario. Let us put the initial position of the system above the singular point $M_{01}$:
\begin{equation}\label{IC1}
\mathbf{I}=[0.9,0.5,1.3,0.5].
\end{equation}
In this case we obtain the phase diagrams (Fig.\ref{ris9} -- \ref{ris11}) which are fundamentally distinct from the considered above: the phase trajectories bypass the singular points of type ``saddle/focus'' and asymptotically press to the boundaries of the prohibited range.
\begin{center}
\includegraphics[width=18cm,height=6cm]{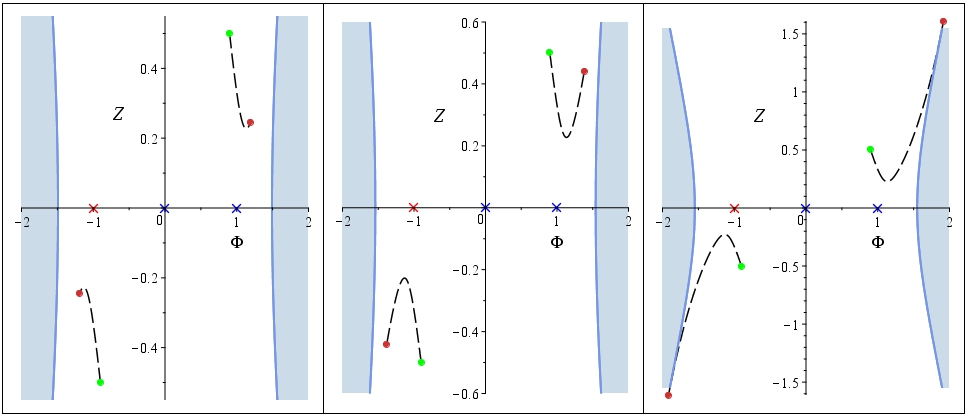}
\figcaption{\label{ris9} The cosmological evolution of the scalar doublet with parameters
 (\ref{param0}) and initial conditions \eqref{IC1} in the ``classical''
 plane $\Sigma_\Phi\equiv\{\Phi,Z\}$. The phase diagrams correspond to the time instants (from left to right): $\tau=1;2;2.2489$.
}
\end{center}

\begin{center}
\includegraphics[width=18cm,height=6cm]{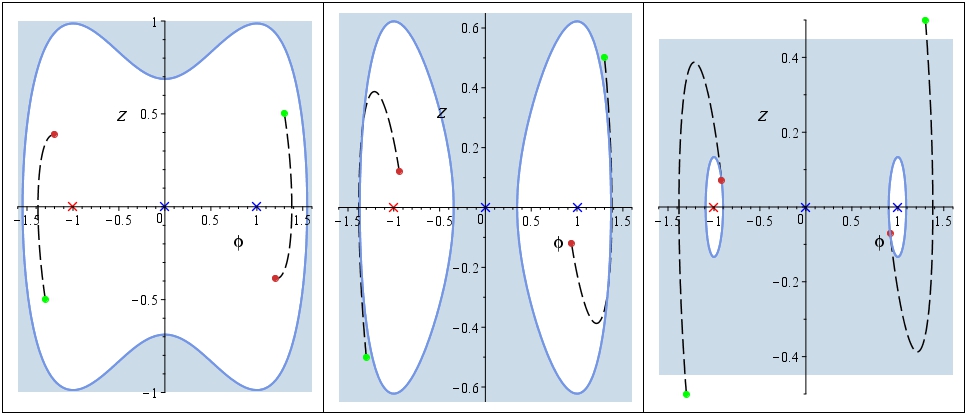}
\figcaption{\label{ris10} The cosmological evolution of the scalar doublet with parameters
 (\ref{param0}) and initial conditions \eqref{IC1} in the phantom plane $\Sigma_\varphi\equiv\{\varphi,z\}$. The phantom diagrams correspond to the time instants (from left to right: $\tau=1;2;2.2489$.
}
\end{center}
\begin{center}
\includegraphics[width=18cm,height=6cm]{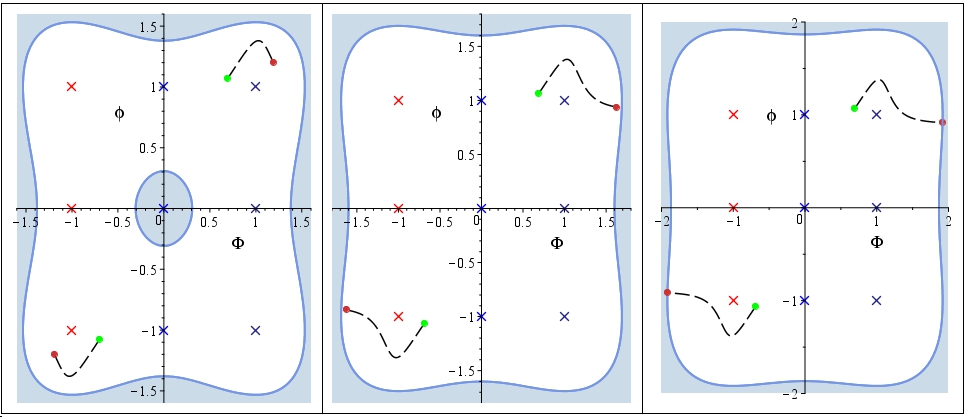}
\figcaption{\label{ris11} The cosmological evolution of the scalar doublet with parameters
 (\ref{param0}) and initial conditions \eqref{IC1} in the plane of potentials $\{\Phi,\varphi\}$. The phase diagrams correspond to the time instants (from left to right): $\tau=1;2;2.2489$.
}
\end{center}
Further, to simultaneously show different-scale and different-sign values we use the Author's one-one continuously differentiable function $\mathrm{Lig}(x)$ (see \cite{Part1}):
\[\mathrm{Lig}(x)\equiv \mathrm{sgn}(x)\log_{10}(1+|x|),\]
so that
\[\mathrm{Lig}(x) \approx \left\{%
\begin{array}{ll}%
x, &  |x|\to 0;\\
\mathrm{sgn}(x)\log_{10}|x|, & |x|\to\infty.
\end{array}
\right.
\]
The Fig. \ref{ris12} shows the evolution of the physical cosmological model with parameters \eqref{param0} under initial conditions \eqref{IC1}:  dimensionless effective energy $\mathcal{E}_m$ (\ref{E_m}),  dimensionless effective pressure $p_m$ (\ref{p_m}) and  invariant cosmological acceleration $\Omega$.
\begin{center}
\includegraphics[width=18cm,height=6cm]{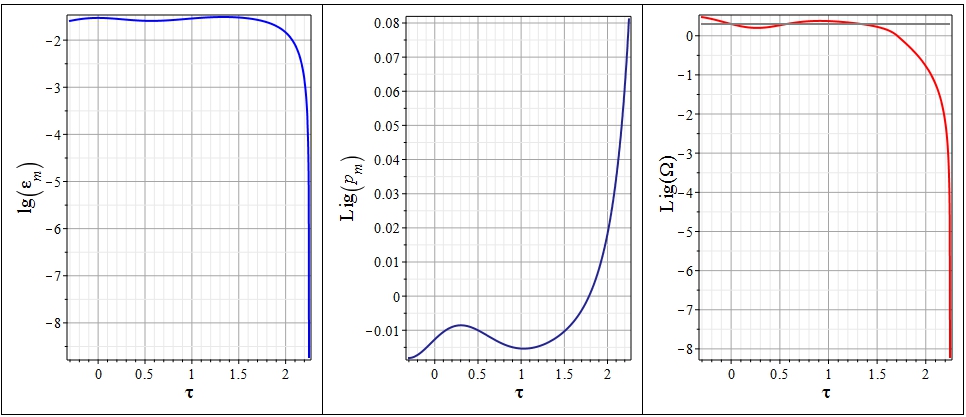}
\figcaption{\label{ris12} The cosmological evolution of the physical characteristics of the cosmological model with parameters
 (\ref{param0}) and initial conditions  \eqref{IC1}. From left to right it is:  dimensionless effective energy  $\mathrm{log10}(\mathcal{E})$ (\ref{E_m});  dimensionless effective pressure $\mathrm{Lig}(p_m)$ (\ref{p_m});  invariant cosmological acceleration $\mathrm{Lig}(\Omega)$ (\ref{Omega}).
The grey horizontal line on the last graph corresponds to the value $\Omega=1$, i.e., inflation.
 }
\end{center}
\begin{multicols}{2}

We see that the effective energy rapidly tends to null, the pressure becomes positive and invariant cosmological acceleration tends to infinitely big negative values. Thus, the full and abrupt halt of the cosmological acceleration happens and the Universe becomes the Euclidian one. The mentioned examples show how diverse the behavior of the cosmological model can be depending on its initial position with respect to the singular points. Let us notice that the results shown on Fig. \ref{ris12} describe the case of transition of the inflationary Universe to the Euclidian one.  As we can see, there is no Big rip problem in our model.

\subsubsection{The Influence of the Model Parameters' Value}
Let us clarify how the absolute values of the parameters $\alpha_m$ and $\beta_m$ influence on the parameters of the model at conservation of the signs of all parameters. Let us consider the following case as an example:
\begin{equation}\label{param0a}
P=[10,10,1,1,1,0].
\end{equation}
The map of singular points in this case coincide with  on Fig. \ref{ris3} provided there is a following change in the coordinates of the points $1\to1/\sqrt{10}$.
%%%%%%%%%%%%%%%%%%%%%%%%%%%%%%%%%%%%%%%%%%%%%%%%%

Fig. \ref{ris13} --\ref{ris15} represents the results of numerical simulation of the dynamic system \eqref{Dyn_sys} for the parameters of the model \eqref{param0} and initial conditions \eqref{IC0a}, corresponding to the initial values of the potentials:
\begin{equation}\label{IC0a}
\mathbf{I}=[0.3,0,0.01,0].
\end{equation}

\end{multicols}
\ruleup
\begin{center}
\includegraphics[width=18cm,height=6cm]{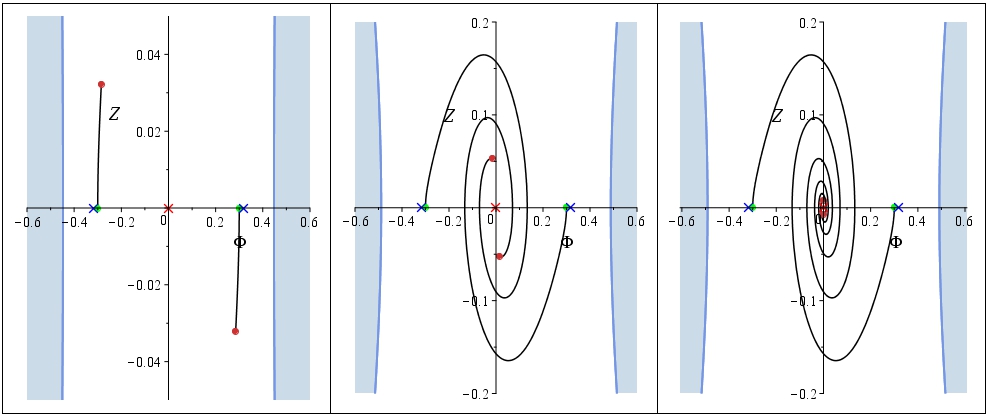}
\figcaption{\label{ris13} The cosmological evolution of the scalar doublet with parameters
 (\ref{param0}) and initial conditions \eqref{IC0} in the ``classical'' plane $\Sigma_\Phi\equiv\{\Phi,Z\}$. The phase diagrams correspond to the time instants (from left to right): $\tau=5;10;20$. The prohibited ranges correspond to the final time instant.
}
\end{center}

\begin{center}
\includegraphics[width=18cm,height=6cm]{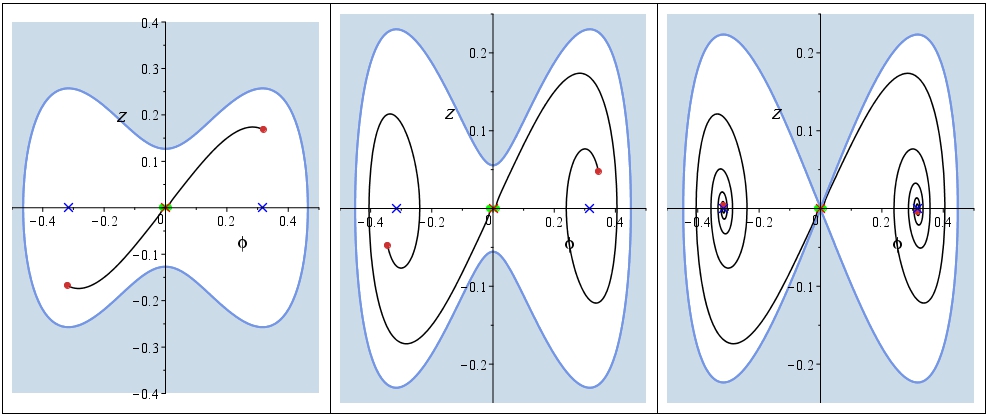}
\figcaption{\label{ris14} The cosmological evolution of the scalar doublet with parameters
 (\ref{param0}) and initial conditions \eqref{IC0} in the phantom plane $\Sigma_\varphi\equiv\{\varphi,z\}$. The phase diagrams correspond to the time instants (from left to right): $\tau=5;10;20$.
}
\end{center}
\begin{center}
\includegraphics[width=18cm,height=6cm]{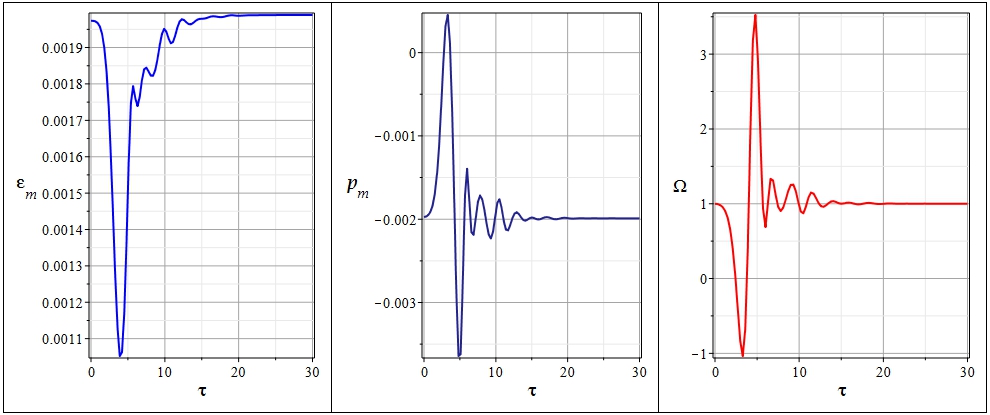}
\figcaption{\label{ris15} The cosmological evolution of physical characteristics of the cosmological model with parameters
 (\ref{param0}) and initial conditions \eqref{IC0}. From left to right it is:  dimensionless effective energy  $\mathcal{E}$ (\ref{E_m});  dimensionless effective pressure $p_m$ (\ref{p_m}); invariant cosmological acceleration $\Omega$ (\ref{Omega}).
 }
\end{center}
%
%%%%%%%%%%%%%%%%%%%%%%%%%%%%%%%%%%%%%%%%%%%%%%%%%
%
\begin{multicols}{2}
It is easily seen that this case does not qualitatively differ from the considered above case with parameters \eqref{param0} (see Fig. \ref{ris5} -- \ref{ris8}). The change of the initial conditions leads to the identical results. The main factor is the accessibility of all singular points.
\subsection{The Bursts of the Cosmological Acceleration\label{Vsplesk}}
As was highlighted in the papers \cite{Ignat_Agaf_Mih_Dima15_3_AST,Ignat_Sasha_G&G,Part1}, the presence of the phantom field in the cosmological model at small values of the phantom field's potential leads to the appearance of the phantom bursts of the super-acceleration, which are characterized by large values of $\Omega$. The Fig. \ref{ris40} shows such bursts.
\begin{center}
\includegraphics[width=8cm]{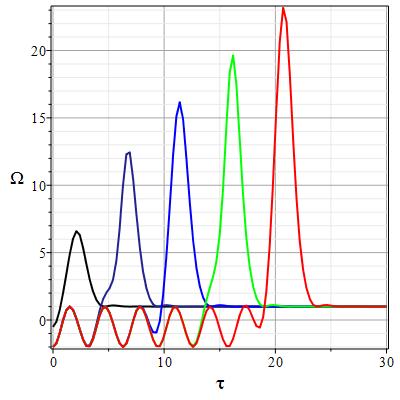}
\figcaption{\label{ris40} The bursts of the cosmological acceleration $\Omega$ for the parameters of the model \eqref{param0} at initial value of the scalar field's potential and its derivative $\Phi(0)=0.1,Z(0)=0$, $z(0)=0$. The black line -- $\varphi(0)=10^{-1}$, dark blue line -- $\varphi(0)=10^{-3}$, cyan line -- $\varphi(0)=10^{-5}$, green line -- $\varphi(0)=10^{-7}$ and red line -- $\varphi(0)=10^{-9}$.
}
\end{center}
As we see, at decrease of the initial value of the phantom field's potential, the burst happens at the later time and its amplitude grows. After the burst, the model comes to the inflation stage.

\section{The Analysis of the Results}
The appendix {\ref{App}} contains the detailed results of numerical simulation of the dynamic system \eqref{Dyn_sys} for all significant cases of the system's parameter sets and initial conditions. In particular, the appendix {\ref{AppA}} shows the results for null value of the cosmological constant and the case of singular points' $M_{01},M_{02}$ inaccessibility; the appendix {\ref{AppB}} considers the case of $M_{10},M_{20}$ singular points' inaccessibility ; the appendix {\ref{AppC}} considers the case of inaccessibility of all singular points except $M_0$; the appendix {\ref{AppD}} investigates the impact of the cosmological constant on the behavior of the dynamic system \eqref{Dyn_sys} for positive values of $\lambda$ while the appendix {\ref{AppF}} investigates it for negative values of $\lambda$.

As the research has shown, the dynamic system \eqref{Dyn_sys} can have 3 types of behavior.\\
\noindent \textbf{A (A Standard Type)}. The phase trajectories in both planes $\Sigma_\Phi$ and $\Sigma_\varphi$ are winding onto the centers/focuses. This case can be split into 3 subcases:\\
\noindent \textbf{$\mathbf{A_1}$}. The phase trajectories in both planes are winding onto null accessible center $M_0$. This case corresponds to parameters \eqref{param1} and initial conditions \eqref{IC2}. Herewith the dynamic system's pressure tends to null, the cosmological acceleration oscillates with the amplitude of order $1$ in the neighbourhood of $\Omega_0=-1/2$, corresponding to the non-relativistic equation of state (Fig. \ref{ris18} -- \ref{ris20}). \\
\noindent \textbf{$\mathbf{A_2}$}. The phase trajectories in the ``classical'' plane are winding onto accessible null center/focus $M_0$, while the phase trajectories in the phantom plane are winding onto accessible non-zero focuses $M_{01},M_{02}$. This case corresponds to parameters \eqref{param0} and initial conditions \eqref{IC0}, The Fig. \ref{ris5} -- \ref{ris8}; parameters \eqref{param0} and initial conditions \eqref{IC0a}, Fig.
\ref{ris13} -- \ref{ris15};  parameters \eqref{param4} and initial conditions \eqref{IC7} ($\lambda>0$), Fig. \ref{ris43} -- \ref{ris45}. The value of the cosmological acceleration in all these cases has a jump $\Omega_{max}>1$, after which the cosmological model comes to inflation regime $\Omega=1$.\\
\noindent \textbf{$\mathbf{A_3}$}. The phase trajectories in the ``classical'' plane are winding onto inaccessible nonzero centers/focuses $M_{10},M_{20}$, while the phase trajectories in the phantom plane are winding onto inaccessible non-zero centers $M_{01},M_{02}$, where the phase trajectories tend to adhere to the boundary of the range with null effective energy in both planes, and fail to do so in the end. As a result, the cosmological acceleration performs 	anharmonic oscillations With greater amplitude around the value $\Omega_0=-1$. This case corresponds to parameters \eqref{param2} and initial conditions \eqref{IC5} (Fig. \ref{ris29} -- \ref{ris31}), and initial conditions \eqref{IC6}, Fig. \ref{ris32} -- \ref{ris34}.\\
\noindent \textbf{B (A Rebound)}. The phase trajectories in the plane $\Sigma_\Phi$ asymptotically tend to line $Z=0$, while the phase trajectories in the plane $\Sigma_\varphi$, tend to infinity by asymptote $\varphi=z,\tau\to\infty$ repulsing from the saddle points $M_{01},M_{02}$ . The cosmological acceleration after the burst comes to the inflation regime -- the model's parameters \eqref{param1}, initial conditions \eqref{IC4}, Fig. \ref{ris25} -- \ref{ris27}, and also -- the model's parameters \eqref{param3}, initial conditions \eqref{IC6}, Fig. \ref{ris37} -- \ref{ris39}.\\
\noindent \textbf{C (An Adherence)}. The phase trajectories in the both planes $\Sigma_\Phi$ and $\Sigma_\varphi$ press to the hypersurface of null effective energy. This happens due to repulsion of the phase trajectory from the saddle point (see, e.g., Fig. \ref{ris9}) or due to attraction to inaccessible center/focus (see e.g. Fig. \ref{ris52}). The effective energy rapidly falls till null, the cosmological acceleration tends to $-\infty$ -- the extremely fast deceleration occurs: parameters \eqref{param0} and initial conditions \eqref{IC1} (Fig. \ref{ris9} -- \ref{ris12}), parameters \eqref{param1} and initial conditions \eqref{IC3} (Fig. \ref{ris21} -- \ref{ris23}), parameters \eqref{param5} and initial conditinos \eqref{IC8} (Fig. \ref{ris48} -- \ref{ris50}), parameters \eqref{param6}and initial conditions \eqref{IC9} (Fig. \ref{ris52} -- \ref{ris54}).
The last two cases correspond to negative values of the cosmological constant. The cosmological acceleration in this case can be negative at certain initial expansion stages.
%%%%%%%%%%%%%%%%%%%%%%%%%%%%%%%%%%%%%%%%%%%%%%%%%%%%%%%%%%%%

Summarizing the results of the research, let us notice that the discovered peculiarities of the dynamic system \eqref{Dyn_sys}, associated, first of all, with the attraction of its phase trajectories to the hypersurfaces of null effective energy, can be laid as a basis of new cosmological scenarios.
Actually, the entry of the cosmological model into stationary orbits with null effective energy at nonzero potentials of the scalar fields and the first derivatives of the scalar doublet allows to treat these orbits as pure vacuum states of these fields, corresponding to null curvature of the Friedmann space-time, i.e. the Euclidian space. This space, as it turns out, can be a not empty one, but containing matter in a form of pair of synchronously oscillating scalar fields which can be considered as virtual vacuum fields. Let us notice that all the considered cases of adherence of the dynamic system to the surfaces of null effective energy correspond to very early stages of the cosmological evolution. The possible instability of this solution which is stationary with respect to the scalar fields' perturbations can become the source of real particles' birth as a result of scalar interactions and not asa result of gravitational instability.
%%%%%%%%%%%%%%%%%%%%%%%%%%%%%%%%%%%%%%%%%%%%%%%%%%%%%%%%%%%%%

\vspace{-1mm}
\centerline{\rule{80mm}{0.1pt}}

%\end{multicols}

\vspace{10mm}

\section{Appendices. The Results of Numerical Simulation\label{App}}

\subsection{Appendix A. The Case of Inaccessibility of the Singular Points $M_{01}$ and $M_{02}$\label{AppA}}
%
%\begin{subequations}
\renewcommand{\thefigure}{A\arabic{figure}}
\renewcommand{\theequation}{A\arabic{equation}}
\begin{equation}\label{param1}
\mathrm{P}=[1,-1,1,-1,1,0].
\end{equation}
\subsubsection{The General Peculiarities of the Phase Space}
In this case the singular points have the following same coordinates as in the previous case \eqref{M1111},
 and invariant characteristics $\sigma^2_i$ are equal to:
 \[\sigma^2_1=-\frac{3}{8};\quad \sigma^2_2=\frac{3}{8};\quad \sigma^2_3=0.\]
Since $\varepsilon\beta_m=1<0$, the singular points $M_{01}$ and $M_{02}$ of the dynamic system are inaccessible. The character of the singular points is shown on Fig. \ref{ris16}.
\begin{center}
\includegraphics[width=6cm]{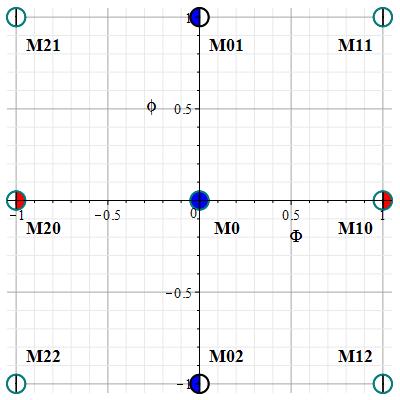}
\figcaption{\label{ris16} The map of singular points in the phase plane $\{\Phi,\varphi\}$ at the model's parameters (\ref{param1}).
}
\end{center}
The Fig. \ref{ris17} illustrates the dependency of the boundaries of the prohibited ranges of the phase space in the projections $\Sigma_\Phi$ and $\Sigma_\varphi$ on the values of dual potentials.
\begin{flushleft}
\includegraphics[width=8.5cm]{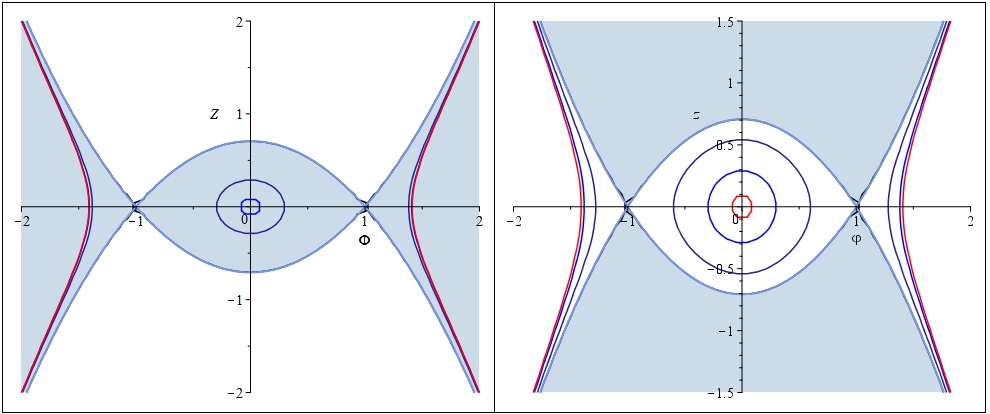}
\figcaption{\label{ris17} The dependency of the prohibited range (marked with a cyan color) in the plane $\Sigma_\Phi$ (on the left) and in the plane $\Sigma_\varphi$ (on the right) from the values of dual potentials at model's parameters (\ref{param1}). On the left:  black line $\varphi=1,z=0$; dark blue line -- $\varphi=0.3,z=0$; cyan line -- $\varphi=0.1,z=0$; red line -- $\varphi=0.01,z=0$; inner ranges are prohibited, the figure illustrates the prohibited range at $\varphi=1,z=0$. On the right:  black line $\Phi=1,Z=0$; dark blue line -- $\Phi=0.6,Z=0$; cyan line -- $\Phi=0.3,Z=0$; red line -- $\Phi=0.1,Z=0$; outer ranges are prohibited, the figure illustrates the prohibited range at $\Phi=1,Z=0$.
}
\end{flushleft}
\subsubsection{The Phase Trajectories of the Dynamic System}
 The Fig. \ref{ris18} -- \ref{ris20} represents the results of numerical simulation of the dynamic system \eqref{Dyn_sys} for the model's parameters \eqref{param1} and initial conditions \eqref{IC2}
\begin{equation}\label{IC2}
\mathbf{I}=[0.7,0.5,0.1,0].
\end{equation}
\end{multicols}
\renewcommand{\thefigure}{A\arabic{figure}}
\renewcommand{\theequation}{A\arabic{equation}}
\ruleup
\begin{center}
\includegraphics[width=18cm,height=6cm]{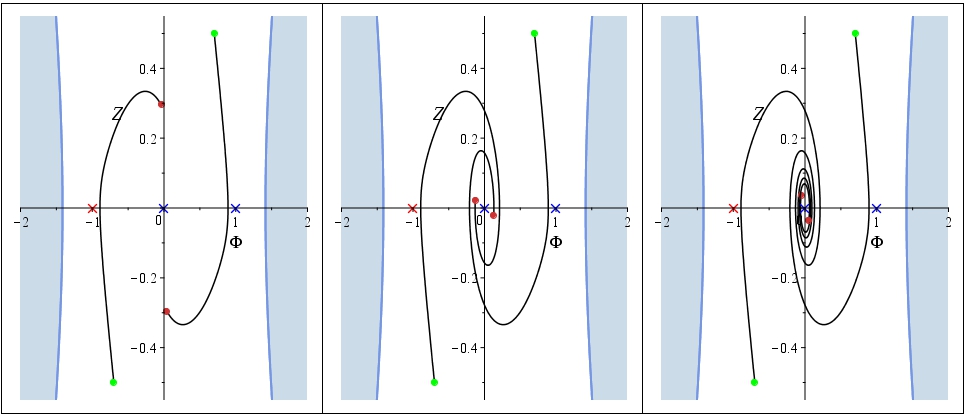}
\figcaption{\label{ris18} The cosmological evolution of the scalar doublet with parameters
 (\ref{param0}) and initial conditions \eqref{IC0} in the ``classical''
 plane $\Sigma_\Phi\equiv\{\Phi,Z\}$. The phase diagrams correspond to the time instants (left to right): $\tau=5;10;20$.
}
\end{center}

\begin{center}
\includegraphics[width=18cm,height=6cm]{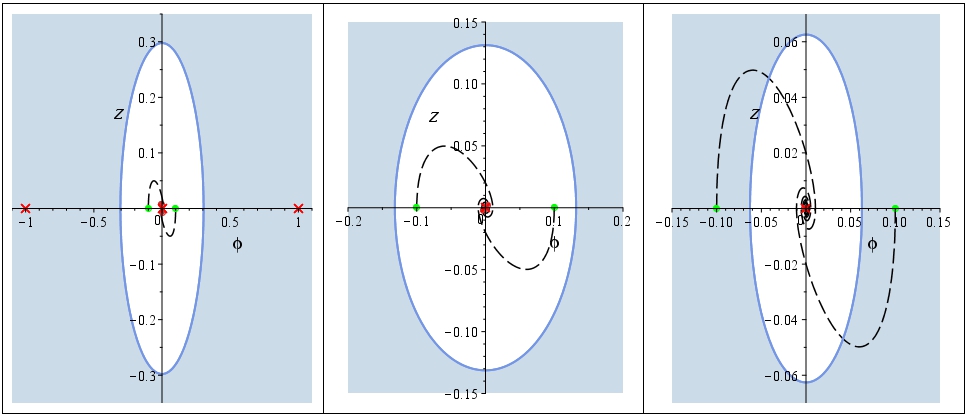}
\figcaption{\label{ris19} The cosmological evolution of the scalar doublet with parameters
 (\ref{param1})and initial conditions \eqref{IC2} in the phantom plane $\Sigma_\varphi\equiv\{\varphi,z\}$. The phase diagrams correspond to the time instants (left to right): $\tau=5;10;20$.
}
\end{center}
The Fig. \ref{ris20} shows the evolution of physical characteristics of the cosmological model with parameters \eqref{param0}: dimensionless effective energy $\mathcal{E}_m$ (\ref{E_m}), dimensionless effective pressure $p_m$ (\ref{p_m}) and invariant cosmological acceleration $\Omega$ (see \cite{Part1})

\begin{center}
\includegraphics[width=18cm,height=6cm]{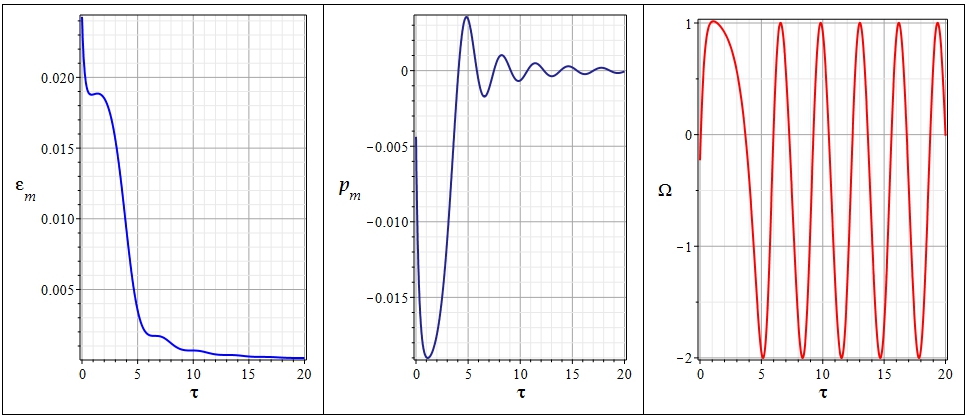}
\figcaption{\label{ris20} The cosmological evolution of the physical characteristics of the cosmological model with parameters
 (\ref{param0}) and initial conditions \eqref{IC2}. Left to right: dimensionless effective energy  $\mathrm{log10}(\mathcal{E})$ (\ref{E_m}); dimensionless effective pressure $\mathrm{Lig}(p_m)$ (\ref{p_m}); invariant cosmological acceleration $\mathrm{Lig}(\Omega)$ (\ref{Omega}).
 The grey horizontal line on the last graph corresponds to the value $\Omega=1$, i.e. to the inflation.
 }
\end{center}
\begin{multicols}{2}
Thus, the phase trajectories of the dynamic system in both planes $\Sigma_\Phi$ and $\Sigma_\varphi$ at initial conditions \eqref{IC2} are winding onto null center $M_0$. The effective energy and the system's pressure herewith tend to null while the invariant cosmological acceleration oscillates around the value $\Omega=-\frac{1}{2}$, corresponding to macroscopic nonrelativistic equation of state
$\varkappa=0$. Exactly the same oscillations appear in the model with classical scalar field (see \cite{Ignat_Sam,Ignat_Sam_Ignat,Ignat_Sam2})\footnote{Let us notice that mentioned oscillations are microscopic ones with the oscillation period equal to the Compton time $1/m$.}.

 The Fig. \ref{ris21} --\ref{ris23} illustrates the results of numerical simulation of the dynamic system \eqref{Dyn_sys} for the model's parameters \eqref{param1} and initial conditions \eqref{IC3}
\begin{equation}\label{IC3}
\mathbf{I}=[0.7,0.5,0.9,0].
\end{equation}
\end{multicols}
\ruleup
\begin{center}
\includegraphics[width=18cm,height=6cm]{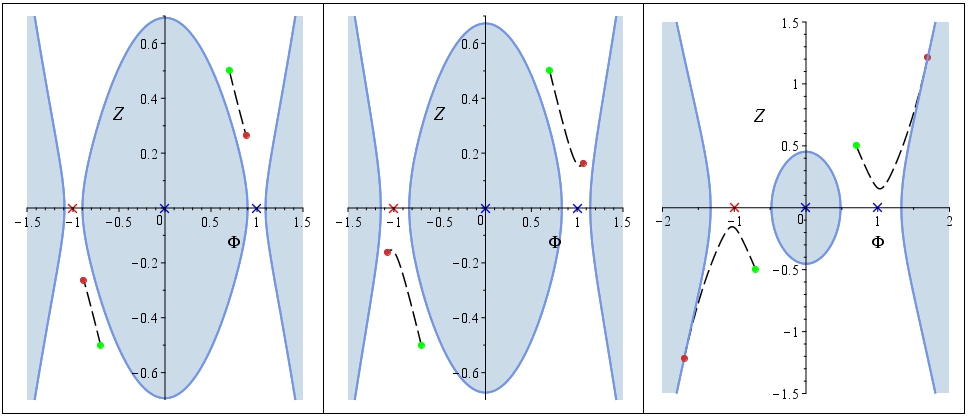}
\figcaption{\label{ris21} The cosmological evolution of the scalar doublet with parameters
 (\ref{param1}) and initial conditions \eqref{IC3} in the ``classical''
 plane $\Sigma_\Phi\equiv\{\Phi,Z\}$. The phase diagrams correspond to the time instants (left to right): $\tau=0.5;1.5;2.99$.
}
\end{center}

\begin{center}
\includegraphics[width=18cm,height=6cm]{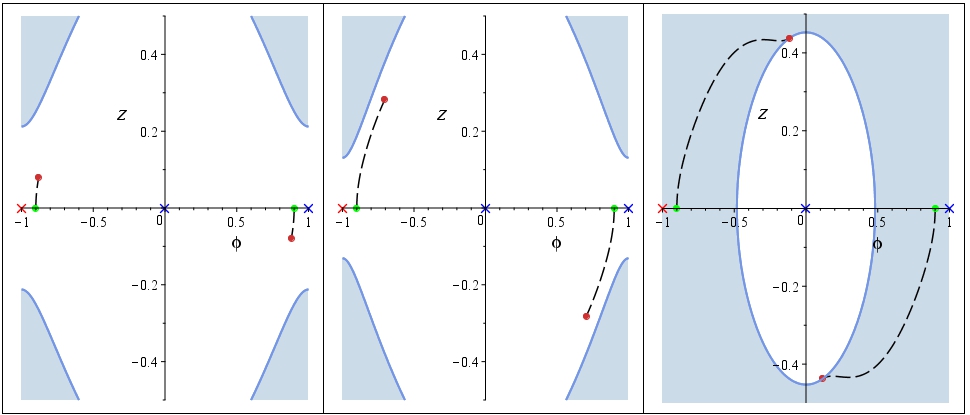}
\figcaption{\label{ris22} The cosmological evolution of the scalar doublet with parameters
 (\ref{param1}) and initial conditions \eqref{IC3} in the phantom plane $\Sigma_\varphi\equiv\{\varphi,z\}$. The phase diagrams correspond to the time instants (left to right): $\tau=0.5;1.5;2.99$.
}
\end{center}

The Fig. \ref{ris23} shows the evolution of the physical characteristics of the cosmological model with parameters \eqref{param1}.

\begin{center}
\includegraphics[width=18cm,height=6cm]{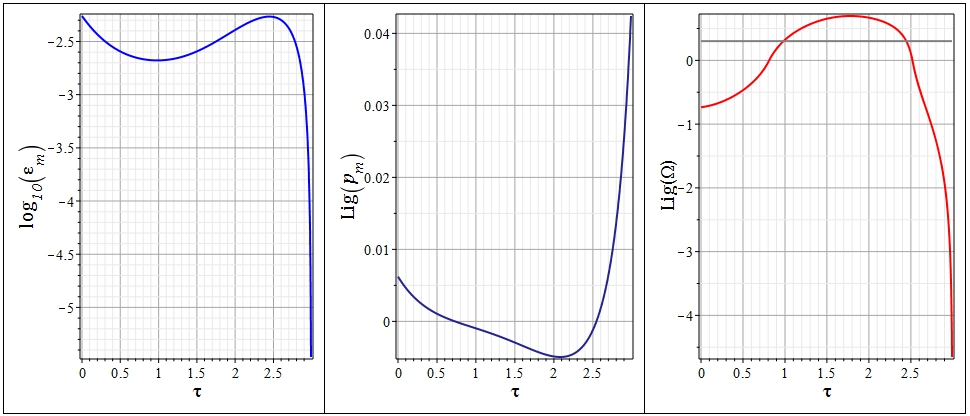}
\figcaption{\label{ris23} The cosmological evolution of physical characteristics of the cosmological model with parameters
 (\ref{param1}) and initial conditions \eqref{IC3}. Left to right: dimensionless effective energy  $\mathrm{log10}(\mathcal{E})$ (\ref{E_m}); dimensionless effective pressure $\mathrm{Lig}(p_m)$ (\ref{p_m}); invariant cosmological acceleration $\mathrm{Lig}(\Omega)$ (\ref{Omega}).
 The grey horizontal line on the last graph corresponds to the value $\Omega=1$, i.e. to the inflation.
 }
\end{center}

\begin{multicols}{2}
\renewcommand{\thefigure}{A\arabic{figure}}
\renewcommand{\theequation}{A\arabic{equation}}
The Fig. \ref{ris24} -- \ref{ris26} represents the results of numerical simulation of the dynamic system \eqref{Dyn_sys} for the model's parameters \eqref{param1} and initial conditions \eqref{IC4}
\begin{equation}\label{IC4}
\mathbf{I}=[0.7,0.5,1.5,0],
\end{equation}
when the dynamic system starts from the position above the singular point $M_{01}$. This case can be called a ``rebound'' -- the system rebounds from the prohibited range in the phantom plane and rapidly increases the potential and kinetic energies of the phantom field, entering the inflation acceleration mode.
\end{multicols}
\ruleup
\begin{center}
\includegraphics[width=18cm,height=6cm]{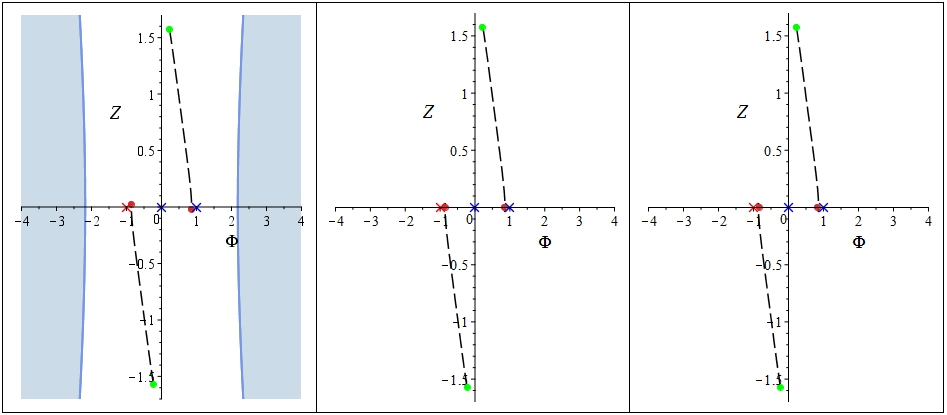}
\figcaption{\label{ris24} The cosmological evolution of the scalar doublet with parameters
 (\ref{param1}) and initial conditions \eqref{IC4} in the ``classical''
 plane $\Sigma_\Phi\equiv\{\Phi,Z\}$. The phase diagrams correspond to the time instants (left to right): $\tau=1;5;8.79$.
}
\end{center}

\begin{center}
\includegraphics[width=18cm,height=6cm]{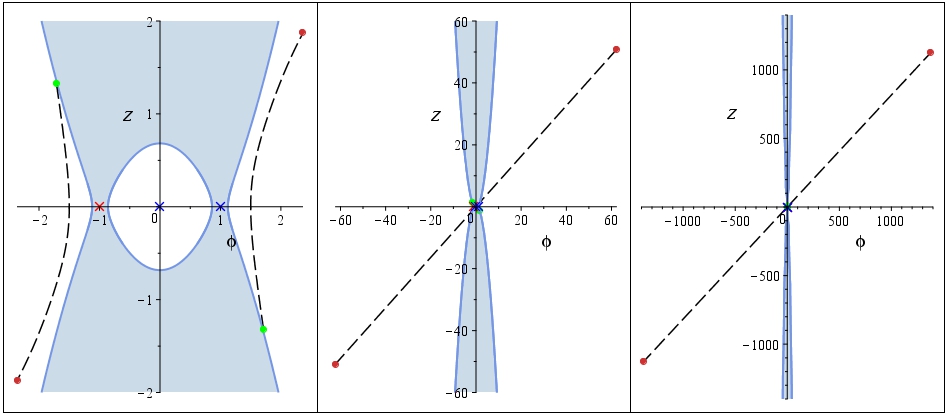}
\figcaption{\label{ris25} The cosmological evolution of the scalar doublet with parameters
 (\ref{param1}) and initial conditions \eqref{IC4} in the phantom plane $\Sigma_\varphi\equiv\{\varphi,z\}$. The phase diagrams correspond to the time instants (left to right): $\tau=1;5;8.79$.
}
\end{center}
\begin{center}
\includegraphics[width=18cm,height=6cm]{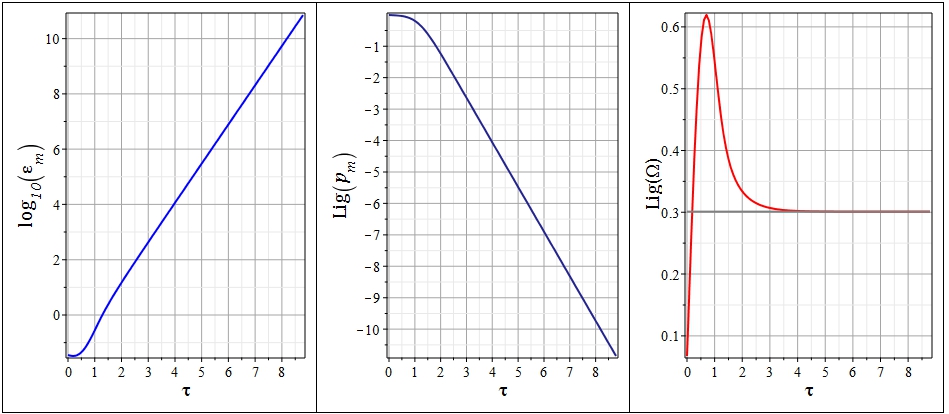}
\figcaption{\label{ris26} The cosmological evolution of physical characteristics of the cosmological model with parameters
 (\ref{param1}) and initial conditions \eqref{IC4}.
 }
\end{center}
The Fig. \ref{ris26} shows the evolution of the physical characteristics of the cosmological model with parameters \eqref{param1}.
%%%%%%%%%%%%%%%%%%%%%%%%%%%%%%%%%%%%%%%%%%%%%%%%%%%%%%%%%%%%%%%%%%%%%%%%%%%%%%%%%%%%%%%%%%%%%%%%%%%%%%%%%%%%%%%%%%%%%
%%%%%%%%%%%%%%%%%%%%%%%%%%%

\begin{multicols}{2}

\subsection{Appendix B. The Case of Inaccessibility of the Singular Points $M_{10}$ and $M_{20}$\label{AppB}}
\renewcommand{\thefigure}{B\arabic{figure}}
\renewcommand{\theequation}{B\arabic{equation}}
\begin{equation}\label{param2}
\mathrm{P}=[-1,1,-1,1,1,0].
\end{equation}
\subsubsection{The General Properties of the Phase Space}
In this case the singular points have the following same coordinates as in the previous case \eqref{M1111},
 and invariant characteristics $\sigma^2_i$ are equal to:
 \[\sigma^2_1=\frac{3}{8};\quad \sigma^2_2=-\frac{3}{8};\quad \sigma^2_3=0.\]
Since it is $\varepsilon\beta_m=1<0$, the singular points $M_{10}$ and $M_{20}$ of the dynamic system are inaccessible. The character of the singular points is presented on the Fig. \ref{ris27}.
\begin{center}
\includegraphics[width=6cm]{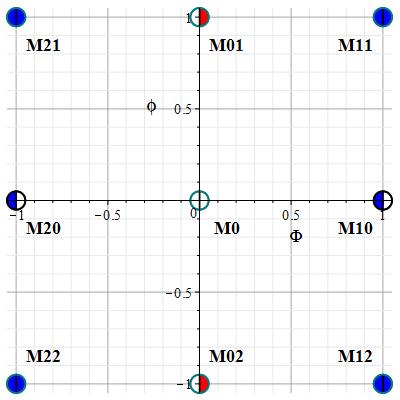}
\figcaption{\label{ris27} The map of singular points in the phase plane $\{\Phi,\varphi\}$ at the model's parameters (\ref{param2}).
}
\end{center}
The Fig. \ref{ris28} shows the dependency of boundaries of the phase space's prohibited ranges in projections $\Sigma_\Phi$ and $\Sigma_\varphi$ on the value of dual potentials.
\begin{flushleft}
\includegraphics[width=8.5cm]{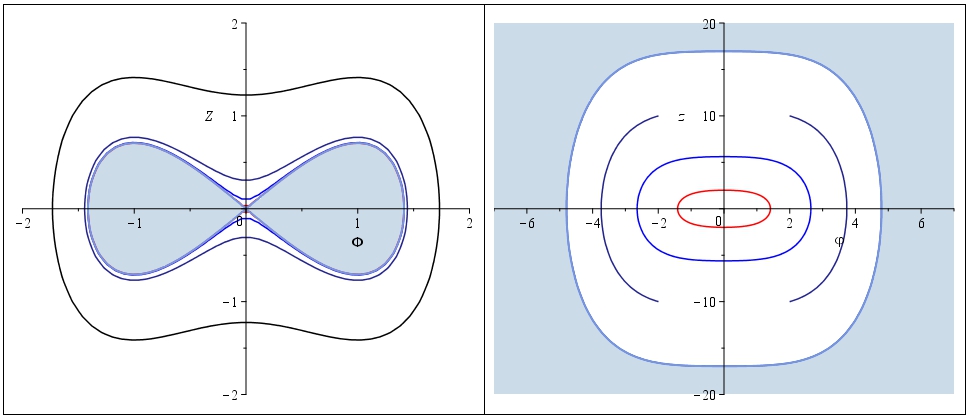}
\figcaption{\label{ris28} The dependency of the prohibited range (highlighted in cyan) in the plane $\Sigma_\Phi$ (on the left) and in the plane $\Sigma_\varphi$ (on the right) on the values of the dual potentials at the model's parameters (\ref{param2}). On the left: black line $\varphi=1,z=0$; dark blue -- $\varphi=0.3,z=0$; cyan -- $\varphi=0.1,z=0$; inner ranges are prohibited, the figure illustrates the prohibited range at $\varphi=0.01,z=0$. On the right: dark blue line $\Phi=4,Z=0$; cyan -- $\Phi=3,Z=0$; red -- $\Phi=0.1,Z=0$; the outer ranges are prohibited \, the figure illustrates the prohibited range at $\Phi=5,Z=0$.
}
\end{flushleft}
\subsubsection{The Phase Trajectories of the Dynamic System}
The Fig. \ref{ris29} --\ref{ris30} represents the results of numerical simulation of the dynamic system \eqref{Dyn_sys} for the model's parameters  \eqref{param2} and initial conditions \eqref{IC5}
\begin{equation}\label{IC5}
\mathbf{I}=[0.5,0.5,0.5,0].
\end{equation}
\end{multicols}
\renewcommand{\thefigure}{B\arabic{figure}}
\renewcommand{\theequation}{B\arabic{equation}}
\ruleup
\begin{center}
\includegraphics[width=18cm,height=6cm]{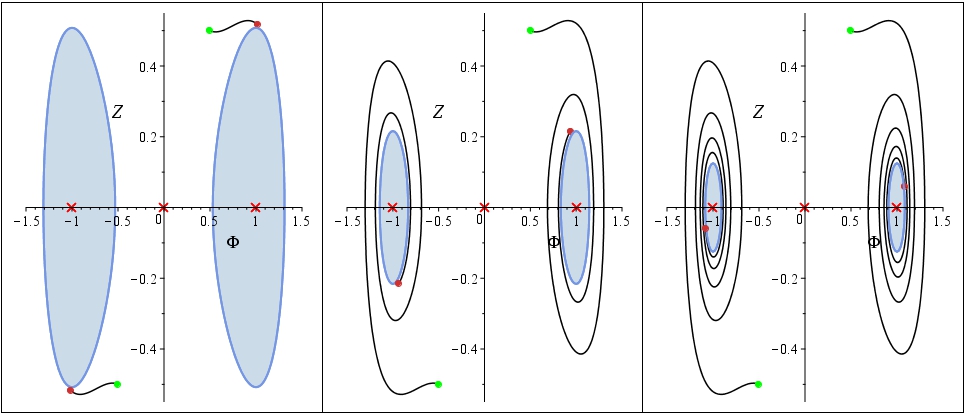}
\figcaption{\label{ris29} The cosmological evolution of the scalar doublet with parameters
 (\ref{param2}) and initial conditions \eqref{IC5} in the ``classical''
 plane $\Sigma_\Phi\equiv\{\Phi,Z\}$. The phase diagrams correspond to the time instants (left to right): $\tau=5;10;20$.
}
\end{center}

\begin{center}
\includegraphics[width=18cm,height=6cm]{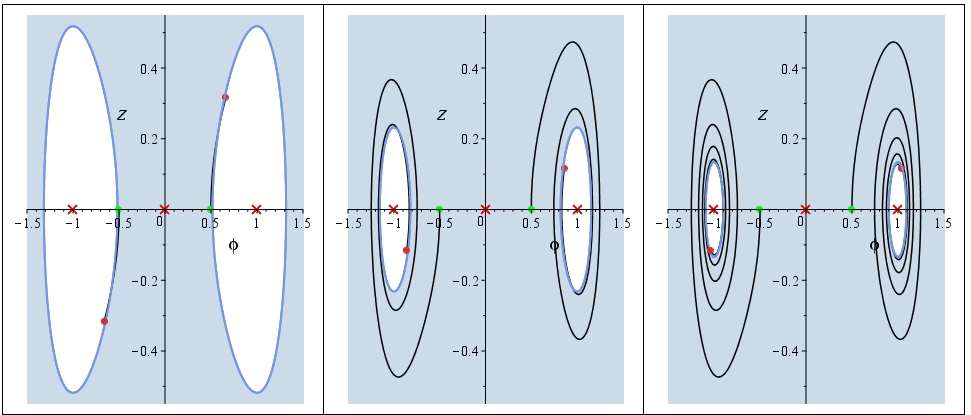}
\figcaption{\label{ris30} The cosmological evolution of the scalar doublet with parameters
 (\ref{param2}) and initial conditions \eqref{IC5} in the phantom plane $\Sigma_\varphi\equiv\{\varphi,z\}$. The phase diagrams correspond to the time instants (left to right): $\tau=5;10;20$.
}
\end{center}
The Fig.\ref{ris31} shows the evolution of the physical characteristics of the cosmological model with parameters \eqref{param2}: dimensionless effective energy $\mathcal{E}_m$ (\ref{E_m}), dimensionless effective pressure $p_m$ (\ref{p_m}) and invariant cosmological acceleration.

\begin{center}
\includegraphics[width=18cm,height=6cm]{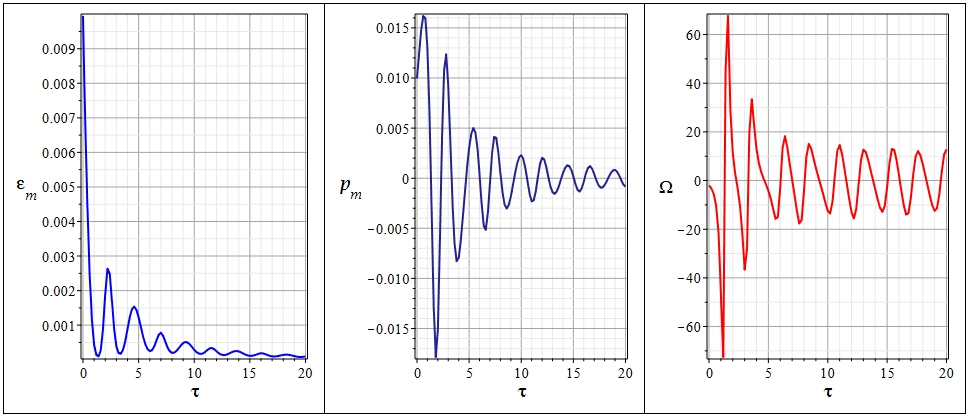}
\figcaption{\label{ris31} The cosmological evolution of the physical characteristics of the cosmological model with parameters
 (\ref{param2}) and initial conditions \eqref{IC5}.
 }
\end{center}
\begin{multicols}{2}
Thus, the phase trajectories of the dynamic system in both planes $\Sigma_\Phi$ and $\Sigma_\varphi$ at initial conditions \eqref{IC5} tend to be winded onto inaccessible centers $M_{10}$ and $M_{20}$ in the plane  $\Sigma_\Phi$ $M_0$ and accessible centers $M_{01}$ and $M_{02}$ in the plane $\Sigma_\varphi$. The trajectories herewith approach closely to the boundary of the prohibited range which leads to the reduction of the effective energy and to the larger values of the acceleration however the regime of full adhesion of the phase trajectory to the prohibited range cannot be set due to different velocities and their change.

The Fig. \ref{ris32} --\ref{ris33} illustrate the results of numerical simulation of the dynamic system \eqref{Dyn_sys} for the model's parameters \eqref{param2} and another group of initial conditions \eqref{IC6}
\begin{equation}\label{IC6}
\mathbf{I}=[1,1,1.5,0].
\end{equation}
\end{multicols}
\ruleup
\begin{center}
\includegraphics[width=18cm,height=6cm]{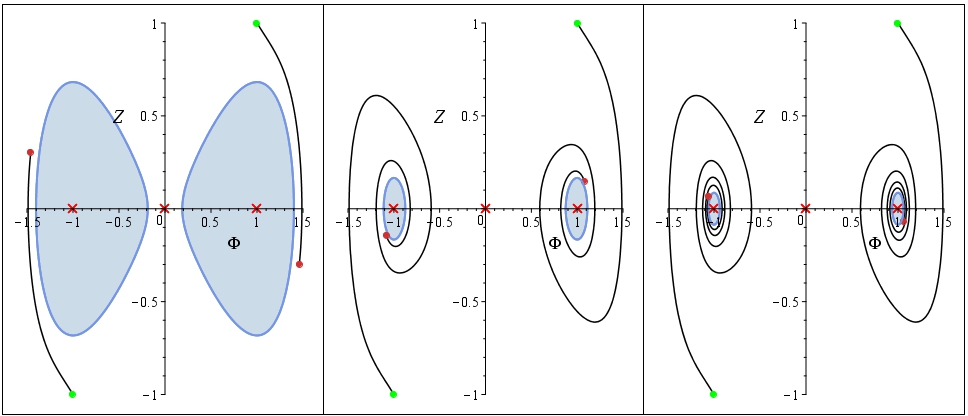}
\figcaption{\label{ris32} The cosmological evolution of the scalar doublet with parameters
 (\ref{param2}) and initial conditions \eqref{IC6} in the ``classical''
 plane $\Sigma_\Phi\equiv\{\Phi,Z\}$. The phase diagrams correspond to the time instants (left to right): $\tau=5;10;20$.
}
\end{center}

\begin{center}
\includegraphics[width=18cm,height=6cm]{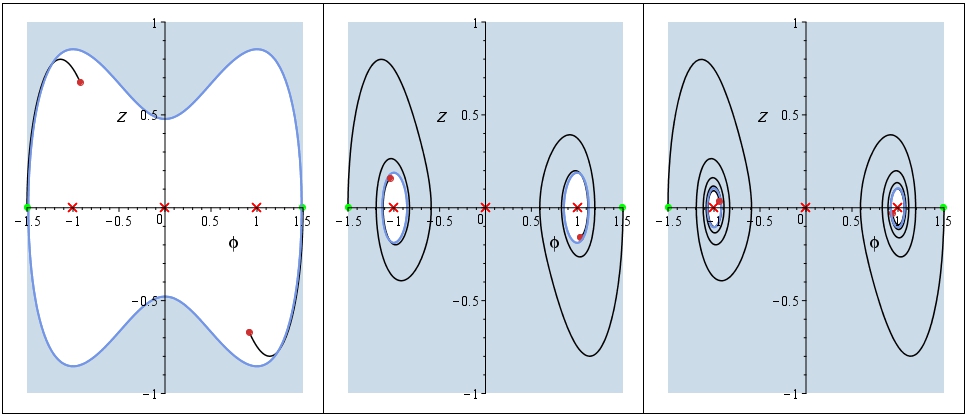}
\figcaption{\label{ris33} The cosmological evolution of the scalar doublet with parameters
 (\ref{param2}) and initial conditions \eqref{IC6} in the phantom plane $\Sigma_\varphi\equiv\{\varphi,z\}$. The phase diagrams correspond to the time instants (left to right): $\tau=5;10;20$.
}
\end{center}
The Fig. \ref{ris34} shows the evolution of the physical characteristics of the cosmological model with parameters \eqref{param2}: dimensionless effective energy $\mathcal{E}_m$ (\ref{E_m}), dimensionless effective pressure $p_m$ (\ref{p_m}) and invariant cosmological acceleration.
As it can be seen, the behavior of the dynamic system at initial conditions \eqref{IC6} practically does not differ at essentially different initial conditions \eqref{IC5} notwithstanding the geography of the prohibited ranges in these cases is essentially different (compare the Fig.-s \ref{ris29} and \ref{ris32}, The Fig.-s \ref{ris30} and \ref{ris33}). This, apparently, testify to the stability of the model's behavior at parameters \eqref{param2} with respect to the change of initial conditions. The consideration of the behavior of the dynamic system at increase of the absolute values of the model's parameters
 $\alpha_m,\beta_m$ by 10 times, at conservation of their signs $\mathbf{P}=[10,-10,1,-1,1,0]$ leads to qualitatively same results with the only difference the cosmological acceleration's oscillation amplitude decreases sharply and is about $1$ (on Fig. \ref{ris34} it exceeds the value $60$).

\begin{center}
\includegraphics[width=18cm,height=6cm]{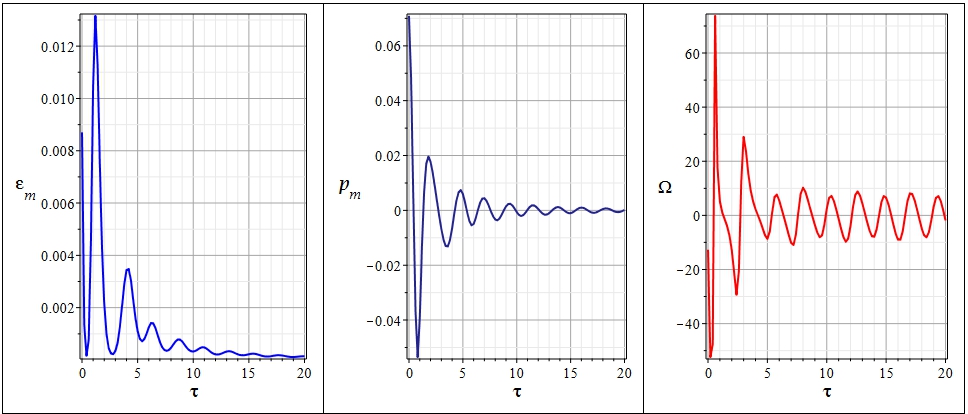}
\figcaption{\label{ris34} The cosmological evolution of the physical characteristics of the cosmological model with parameters
 (\ref{param2}) and initial conditions \eqref{IC6}.
 }
\end{center}
\begin{multicols}{2}

\subsection{Appendix C. The Case of Inaccessibility of All Singular Points Except $M_0$\label{AppC}}
\renewcommand{\theequation}{C\arabic{equation}}
\renewcommand{\thefigure}{Ñ\arabic{figure}}
\begin{equation}\label{param3}
\mathrm{P}=[-1,-1,-1,-1,1,0].
\end{equation}
\subsubsection{The General Properties of the Phase Space}
The singular points have the following same coordinates as in the previous case \eqref{M1111},
 and the invariant characteristics $\sigma^2_i$ are equal to:
 \[\sigma^2_1=-\frac{3}{8};\quad \sigma^2_2=\frac{3}{8};\quad \sigma^2_3=0.\]
Since it is $e\alpha_m=1<0$ and $\varepsilon\beta_m=1<0$, all the singular points of the dynamic system except  $M_0$ are inaccessible. The character of the singular points is represented on the scheme Fig. \ref{ris35}.

\begin{center}
\includegraphics[width=6cm]{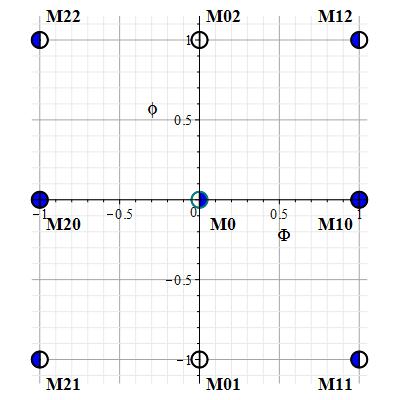}
\figcaption{\label{ris35} The map of singular points in the phase plane $\{\Phi,\varphi\}$ at the model's parameters \eqref{param3}.}
\end{center}

The Fig. \ref{ris36} shows the dependency of the boundaries of the prohibited ranges of the phase space in the projections $\Sigma_\Phi$ and $\Sigma_\varphi$ on the value of dual potentials.
\begin{flushleft}
\includegraphics[width=8.5cm]{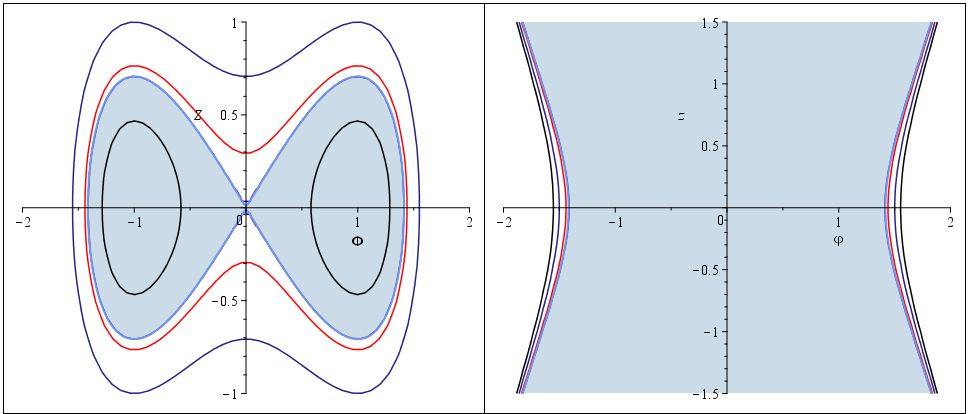}
\figcaption{\label{ris36} The dependency of the prohibited range (highlighted in cyan color) in the plane $\Sigma_\Phi$ (on the left) and in the plane $\Sigma_\varphi$ (on the right) on the values of dual potentials at the model's parameters (\ref{param3}). On the left: black line $\varphi=1.5,z=0$; dark blue line -- $\varphi=13,z=0$; red line -- $\varphi=0.3,z=0$; inner ranges are prohibited, the figure shows the prohibited range at $\varphi=0.01,z=0$. On the right: black line $\Phi=1,Z=0$; dark blue line -- $\Phi=0.6,Z=0$; red line -- $\Phi=0.3,Z=0$; outer ranges are prohibited, the figure shows the proibited range at $\Phi=0.1,Z=0$.
}
\end{flushleft}
%
%%%%%%%%%%%%%%%%%%%%%%%%%%%%%%%%%%%%%%%%%%%%%%%%%%%%%%%%%%%%%%%%%%%%%%%%%%%%%%%%%%%%
\end{multicols}
\renewcommand{\theequation}{C\arabic{equation}}
\renewcommand{\thefigure}{Ñ\arabic{figure}}
\ruleup
\begin{center}
\includegraphics[width=18cm,height=6cm]{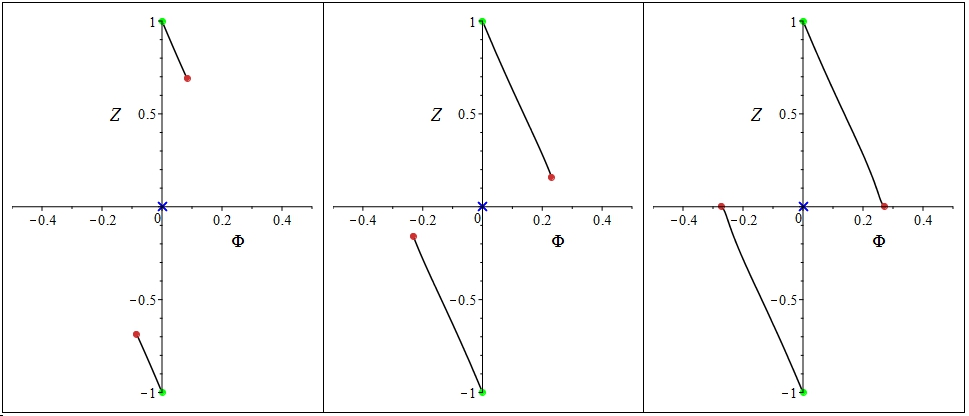}
\figcaption{\label{ris37} The cosmological evolution of the scalar doublet with parameters
 (\ref{param2}) and initial conditions \eqref{IC6} in the ``classical''
 plane $\Sigma_\Phi\equiv\{\Phi,Z\}$. The phase diagrams correspond to the time instants (left to right): $\tau=5;10;20$.
}
\end{center}

\begin{center}
\includegraphics[width=18cm,height=6cm]{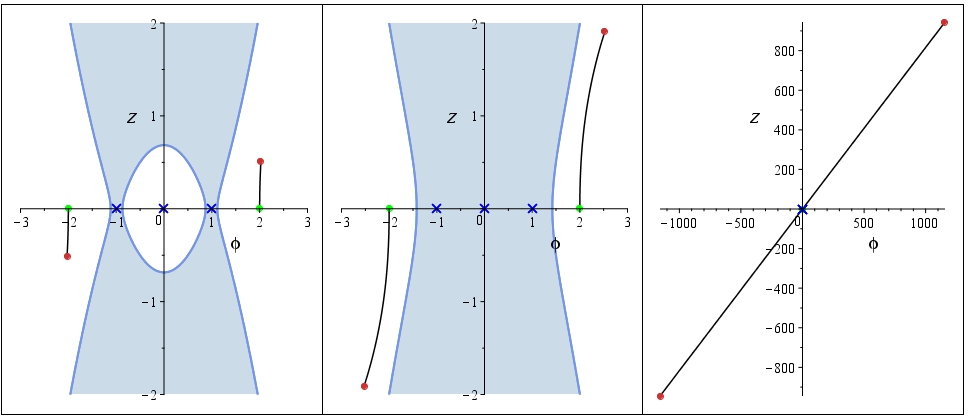}
\figcaption{\label{ris38} The cosmological evolution of the scalar doublet with parameters
 (\ref{param2}) and initial conditions \eqref{IC6} in the phantom plane $\Sigma_\varphi\equiv\{\varphi,z\}$. The phase diagrams correspond to the time instants (left to right): $\tau=5;10;20$.
}
\end{center}

\begin{center}
\includegraphics[width=18cm,height=6cm]{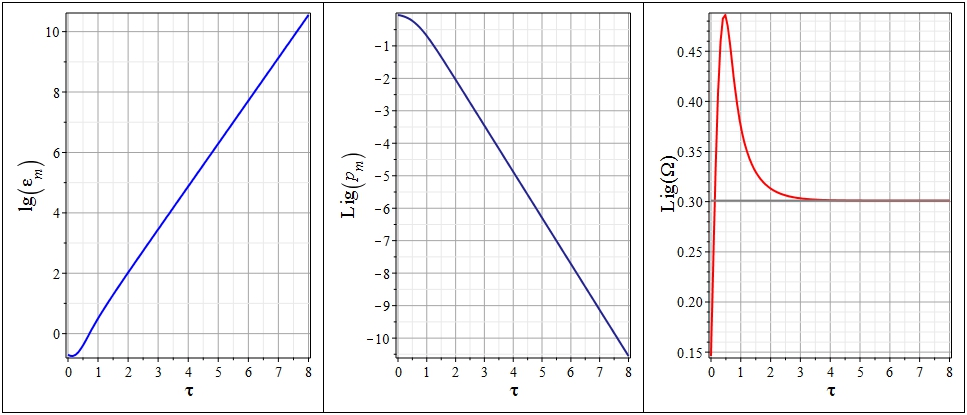}
\figcaption{\label{ris39} The cosmological evolution of the physical characteristics of the cosmological models with parameters
 (\ref{param2}) and initial conditions \eqref{IC6}.
 }
\end{center}
\begin{multicols}{2}
The Fig. \ref{ris39} shows the evolutino of the physical characteristics of the cosmological model with parameters \eqref{param2}: dimensionless effective energy $\mathcal{E}_m$ (\ref{E_m}), dimensionless effective pressure $p_m$ (\ref{p_m}) and invariant cosmological acceleration.
Comparing this case with the case of \eqref{param1}, \eqref{IC4} (the Fig. \ref{ris21} -- \ref{ris23}) we see the qualitative coincidence in behavior of the cosmological models i.e. we obtain the behavior of the model of the ``rebound'' type.
\subsection{Appendix D. The Impact of the Cosmological Constant on the Behavior of the Dynamic System. The Positive Values of the Cosmological Constant\label{AppD}}
\renewcommand{\theequation}{D\arabic{equation}}
\renewcommand{\thefigure}{D\arabic{figure}}
As our researches show, the existence of small positive cosmological constant $\lambda_m\lesssim 0.1$ does not impact significantly on the behavior of the considered dynamic system. Accordingly, we will not discuss in details the results of numerical simulation for this case, limiting ourselves to the single example and certain remarks. Let us first consider the case, equivalent to \eqref{param0a}:
\begin{equation}\label{param4}
\mathrm{P}=[10,10,1,1,1,0.1].
\end{equation}
In this case singular points have the same coordinates as in the case \eqref{M1111},
the invariant characteristics $\sigma^2_i$ are all also positive and all 9 singular points of the dynamic system are accessible. However, the character of these points shown on  Fig. \ref{ris41}, slightly differs from the character if the points on the Fig. \ref{ris3} -- null singular point $M_0$ now is an attractive focus in the plane $\Sigma_\Phi$:
\begin{center}
\includegraphics[width=6cm]{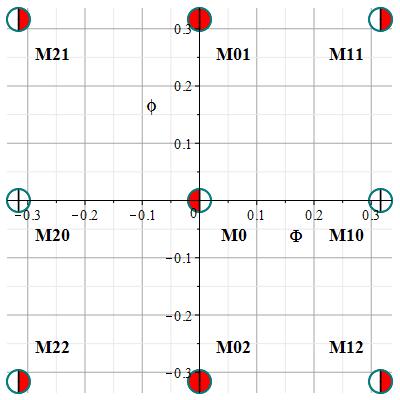}
\figcaption{\label{ris41} The map of singular points in the phase plane $\{\Phi,\varphi\}$ at the model's parameters (\ref{param4}).
}
\end{center}
The boundaries of the prohibited ranges (Fig. \ref{ris42}) behave themselves the same way as on the Fig. \ref{ris4}. The change of the center to the attractive focus practically does not change the situation. Exactly for that reason, the behavior of the model in the case of small values of the cosmological constant $\lambda_m$ practically does not differ from the behavior of the corresponding model with null value of the cosmological constant.
\begin{flushleft}
\includegraphics[width=8.5cm]{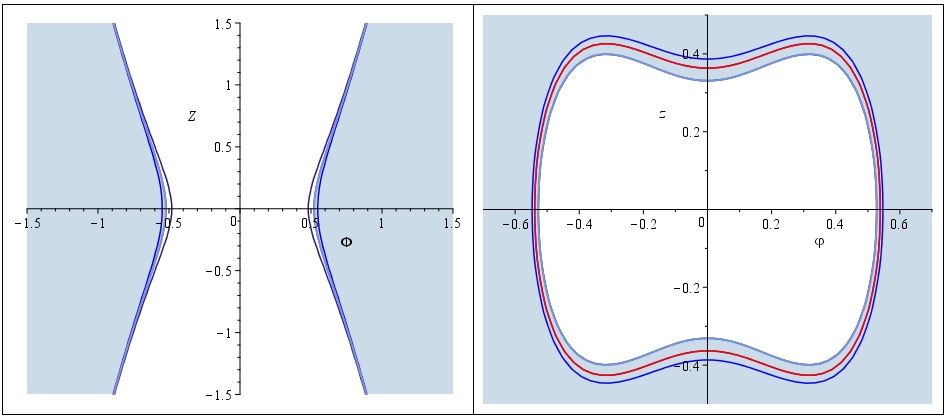}
\figcaption{\label{ris42} The dependency of the prohibited range (highlighted by cyan color) in the plane $\Sigma_\Phi$ (on the left) and in the plane $\Sigma_\varphi$ (on the right) on the values of dual potentials at the model's parameters (\ref{param4}). On the left: black line $\varphi=1,z=0$; dark blue line -- $\varphi=0.5,z=0$; cyan line -- $\varphi=0.3,z=0$; inner ranges are prohibited, the figure shows the prohibited range at $\varphi=0.01,z=0$. On the right: dark blue line -- $\Phi=0.2,Z=0$; red line -- $\Phi=0.4,Z=0$; inner ranges are prohibited, the figure shows the prohibited range at $\Phi=0.1,Z=0$.
}
\end{flushleft}
Since the behavior of the system differs slightly from the case of null value of the cosmological constant, let us bring one example of the research.
The Fig.-s \ref{ris43} --\ref{ris45} show the results of numerical simulation of the dynamic system \eqref{Dyn_sys} for the model's parameter \eqref{param4} and initial conditions \eqref{IC7}
\begin{equation}\label{IC7}
\mathbf{I}=[0,0.1,0.01,0].%
\end{equation}
\end{multicols}
\renewcommand{\theequation}{D\arabic{equation}}
\renewcommand{\thefigure}{D\arabic{figure}}
\ruleup
\begin{center}
\includegraphics[width=18cm,height=6cm]{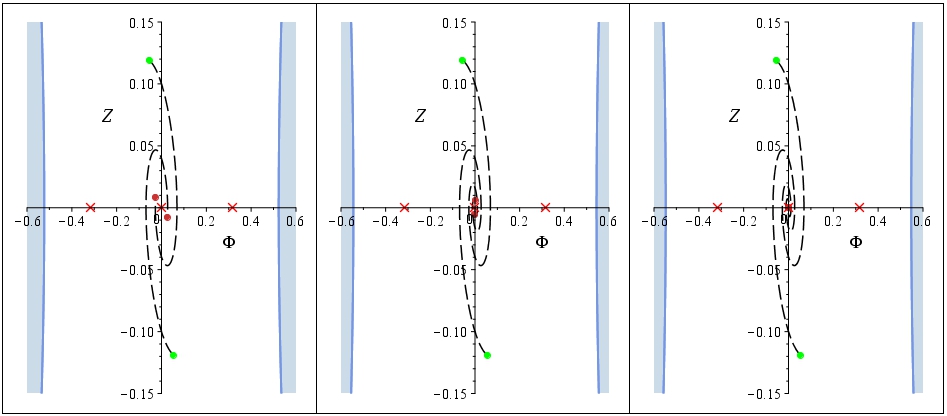}
\figcaption{\label{ris43} The cosmological evolution of the scalar doublet with parameters
 (\ref{param4}) and initial conditions \eqref{IC7} in the ``classical'' plane $\Sigma_\Phi\equiv\{\Phi,Z\}$. The phase diagrams correspond to the time instants (left to right): $\tau=5;10;50$.
}
\end{center}

\begin{center}
\includegraphics[width=18cm,height=6cm]{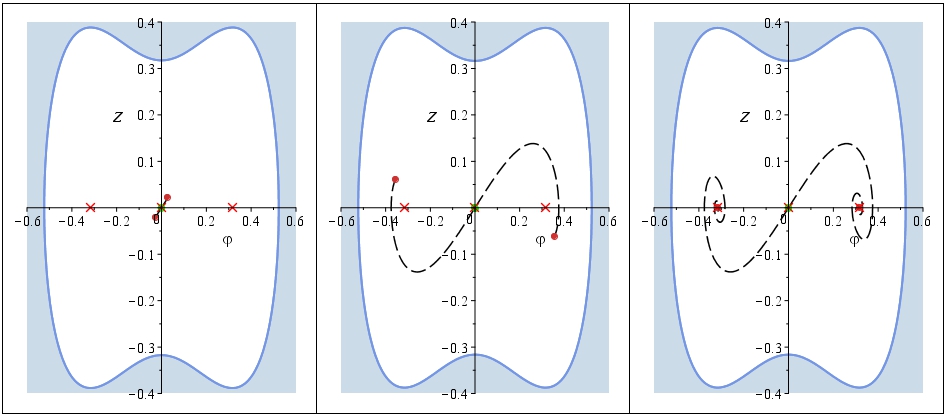}
\figcaption{\label{ris44} The cosmological evolution of the scalar doublet with parameters
 (\ref{param4})and initial conditions \eqref{IC7} in the phantom plane $\Sigma_\varphi\equiv\{\varphi,z\}$. The phase diagrams correspond to the time instants (left to right): $\tau=5;10;50$.
}
\end{center}

\begin{center}
\includegraphics[width=18cm,height=6cm]{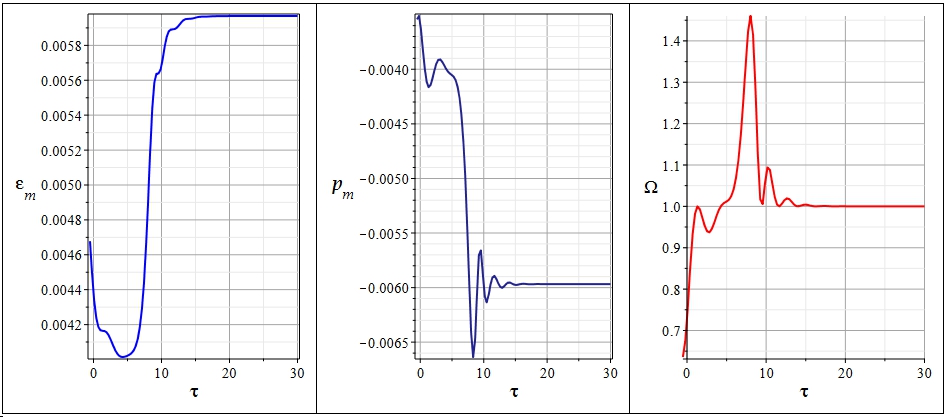}
\figcaption{\label{ris45} The cosmological evolution of the physical characteristics of the cosmological model with parameters
 (\ref{param4}) and initial conditions \eqref{IC7}.
 }
\end{center}
\begin{multicols}{2}
As it can be seen, this case is largely identical to the cases \eqref{param0}, \eqref{IC0} and \eqref{param0a}, \eqref{IC0a}. Let us also notice that the bursts of the cosmological acceleration decrease with growth of the cosmological constant.

%%%%%%%%%%%%%%%%%%%%%%%%%%%%%%%%%%%%%%%%%%%%%%%%%%%%%%%%%%%%%%%%%%%%%%%%%%%%%%%%%%%%%%%%%%%%%%%%%%%%%%
\subsection{Appendix E. The Impact of the Cosmological Constant on the Behavior of the Dynamic System. Negative Values of the Cosmological Constant\label{AppF}}
\renewcommand{\theequation}{E\arabic{equation}}
\renewcommand{\thefigure}{E\arabic{figure}}
Let us consider an ``exotic'' case of negative values of the cosmological constant $\lambda<0$. Let us first of all notice that in this case significant difficulties related to the numerical simulation of the dynamic system \eqref{Dyn_sys} appear. Let us also notice that notwithstanding all the exoticism, the case $\lambda<0$ is of interest to us due to the fact it discovers unique and yet not investigated possibilities for the cosmology. In the first instance these possibilities are associated with the tendency of the dynamic system's ``adhesion'' to hypersurfaces of null effective energy which, in turn, provides a possibility of the appearance of the Euclidian, vacuum stages of the Universe's evolution. Let us consider the case, identical to \eqref{param0a} in details:
\begin{equation}\label{param5}
\mathrm{P}=[10, 10, 1, 1, 1, -0.1].
\end{equation}
In this case there are accessible only 4 singular points of the dynamic system $M_{11},M_{12},M_{21},M_{22}$, which are saddle ones in the plane $\Sigma_\Phi$ and centers in the plane $\Sigma_\varphi$ (Fig. \ref{ris46}).
\begin{center}
\includegraphics[width=6cm]{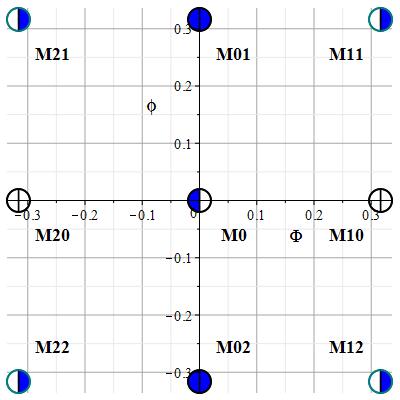}
\figcaption{\label{ris46} The map of singular points in the phase plane $\{\Phi,\varphi\}$ at the model's parameters (\ref{param5}).
}
\end{center}
The boundaries of the prohibited ranges (Fig. \ref{ris47}) in this case behave themselves essentially different from the other cases.
\begin{flushleft}
\includegraphics[width=8.5cm]{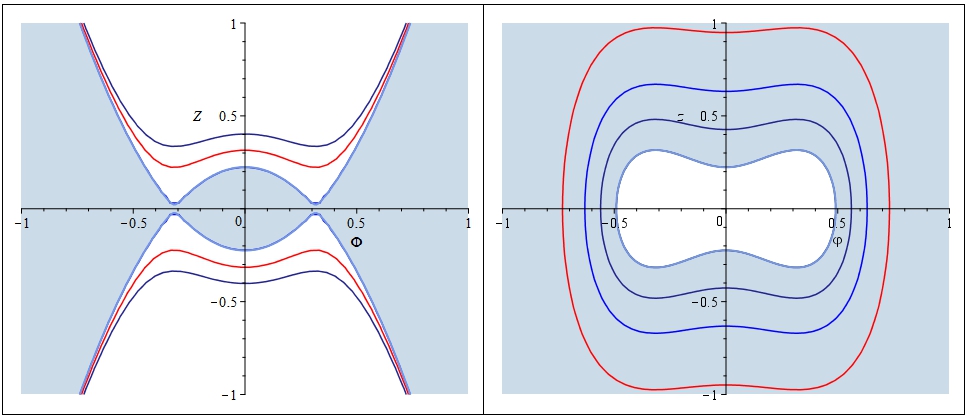}
\figcaption{\label{ris47} The dependency of the prohibited range (highlighted by cyan color) in the plane $\Sigma_\Phi$ (on the left) and in the plane $\Sigma_\varphi$ (on the right) on the values of dual potentials at the model's parameters (\ref{param5}). On the left: black line $\varphi=1,z=0$; dark blue line -- $\varphi=0.5,z=0$; cyan line -- $\varphi=0.3,z=0$; red line -- $\varphi=0.01,z=0$ inner ranges are prohibited, the figure shows the prohibited range at $\varphi=0.3,z=0$. On the right: black line -- $\Phi=1/\sqrt{10},Z=1/\sqrt{10}$, dark blue line -- $\Phi=0.2,Z=0.5$, cyan line -- $\Phi=0.1,Z=0.7$, red line -- $\Phi=0.01,Z=1$; outer ranges are prohibited, the figure shows the prohibited range at $\Phi=1/\sqrt{10},Z=1/\sqrt{10}$.
}
\end{flushleft}
The results of numerical simulation of the dynamic system \eqref{Dyn_sys} at initial conditions are presented below
\begin{equation}\label{IC8}
\mathbf{I}=[0,0.4,0.1,0].%
\end{equation}
\end{multicols}
\renewcommand{\theequation}{E\arabic{equation}}
\renewcommand{\thefigure}{E\arabic{figure}}
\ruleup
\begin{center}
\includegraphics[width=18cm,height=6cm]{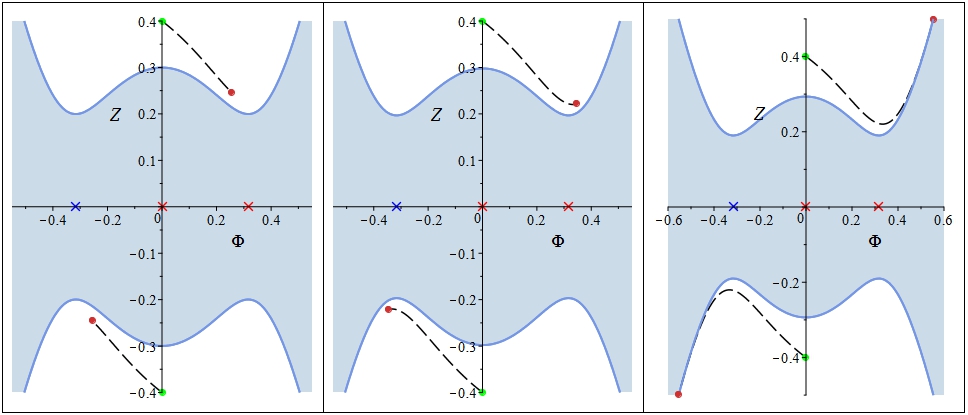}
\figcaption{\label{ris48} The cosmological evolution of the scalar doublet with parameters
 (\ref{param5}) and initial conditions \eqref{IC8} in the ``classical''
 plane $\Sigma_\Phi\equiv\{\Phi,Z\}$.
}
\end{center}

\begin{center}
\includegraphics[width=18cm,height=6cm]{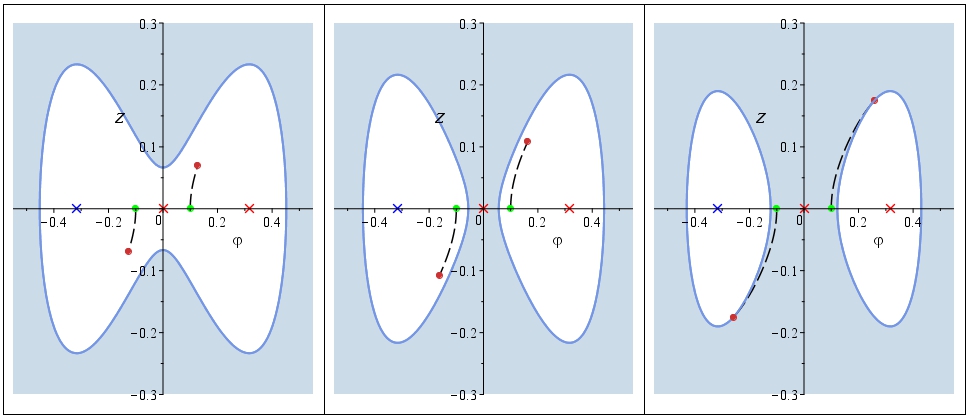}
\figcaption{\label{ris49} The cosmological evolution of the scalar doublet with parameters
 (\ref{param5}) and initial conditions \eqref{IC8} in the phantom plane $\Sigma_\varphi\equiv\{\varphi,z\}$.
}
\end{center}

\begin{center}
\includegraphics[width=18cm,height=6cm]{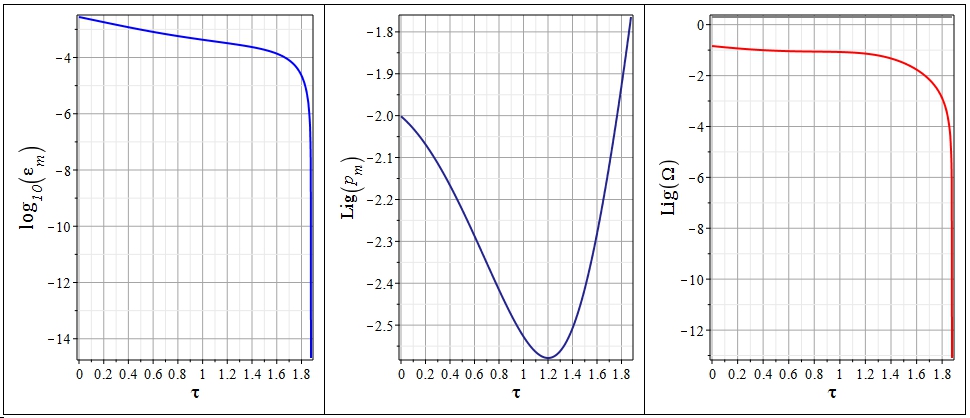}
\figcaption{\label{ris50} The cosmological evolution of the physical characteristics of the cosmological model with parameters
 (\ref{param5}) and initial conditions \eqref{IC8}.
 }
\end{center}
\begin{multicols}{2}
We see that in this case the phase trajectories ``adhere'' to hypersurfaces of null effective energy, as evidenced by the sharp deceleration of the cosmological expansion (the value $\Omega$ grows in negative range to values of order $10^{-14}$!. However, in the considered above case the acceleration at each stage of stays negative which makes the considered cosmological model impractical.

Let us consider a dynamic system with the following parameters:
\renewcommand{\theequation}{E\arabic{equation}}
\renewcommand{\thefigure}{E\arabic{figure}}
\begin{equation}\label{param6}
\mathrm{P}=[-1, 1, -1, 1, 1, -0.1].
\end{equation}
The map of singular points is presented on the Fig. \ref{ris51}
\begin{center}
\includegraphics[width=6cm]{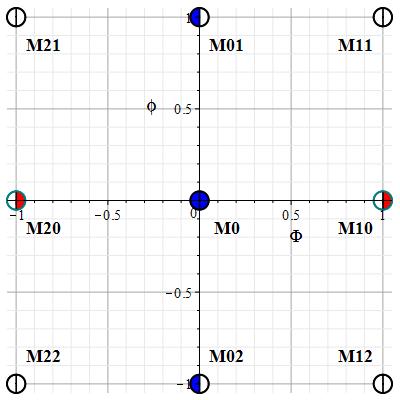}
\figcaption{\label{ris51} The map of singular points in the phase plane $\{\Phi,\varphi\}$ at the model's parameters (\ref{param6}).
}
\end{center}
As it can be seen, only two singular points of the dynamic system are accessible: $M_{10}$ and $M_{20}$, and they are saddle ones in the plane $\Sigma_\Phi$ and attractive focuses in the plane $\Sigma_\varphi$. The Fig. \ref{ris52} -- \ref{ris54} represents the results of numerical simulation of the dynamic system \eqref{Dyn_sys} with parameters \eqref{param6} and initial conditions
\begin{equation}\label{IC9}
\mathbf{I}=[2.5,1,0.5,0.50].%
\end{equation}
\end{multicols}
\ruleup
\begin{center}
\includegraphics[width=18cm,height=6cm]{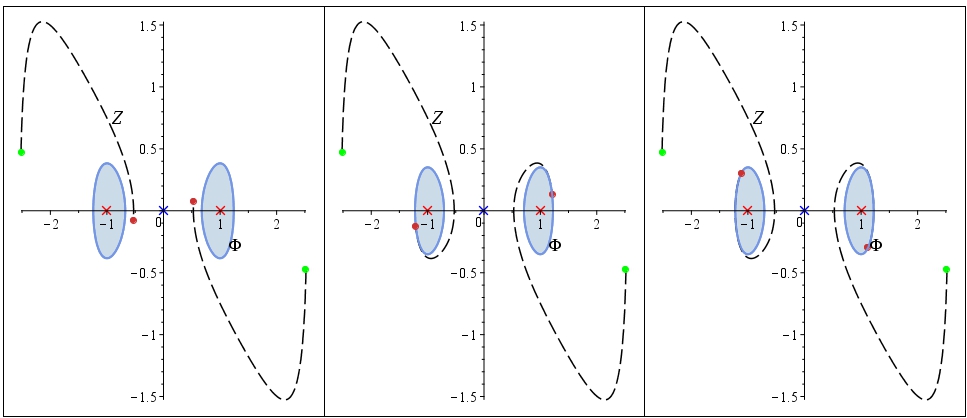}
\figcaption{\label{ris52} The cosmological evolution of the scalar doublet with parameters
 (\ref{param6}) and initial conditions \eqref{IC9} in the ``classical''
 plane $\Sigma_\Phi\equiv\{\Phi,Z\}$.
}
\end{center}

\begin{center}
\includegraphics[width=18cm,height=6cm]{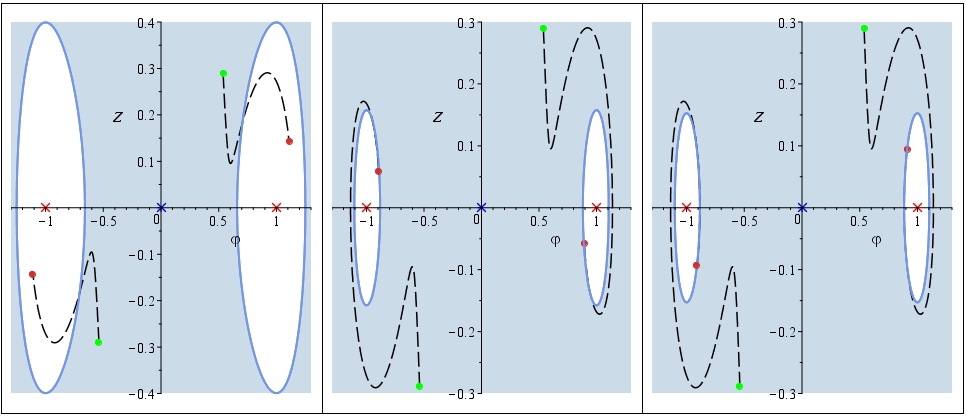}
\figcaption{\label{ris53} The cosmological evolution of the scalar doublet with parameters
 (\ref{param5}) and initial conditions \eqref{IC8} in the phantom plane $\Sigma_\varphi\equiv\{\varphi,z\}$.
}
\end{center}

\begin{center}
\includegraphics[width=18cm,height=6cm]{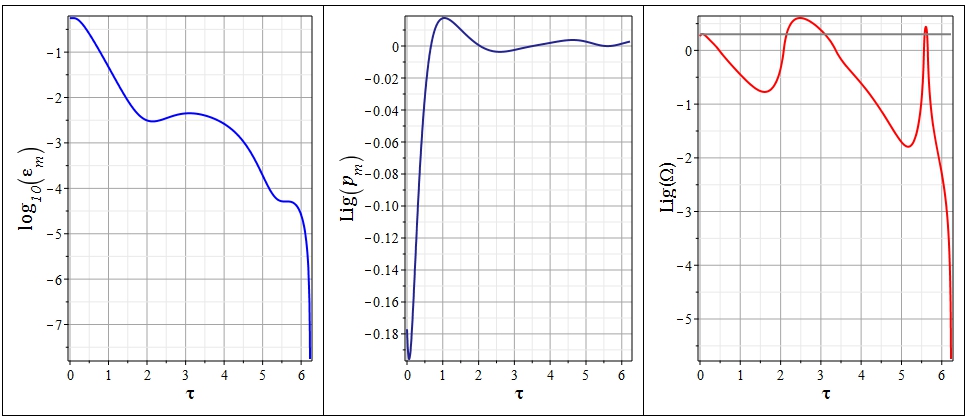}
\figcaption{\label{ris54} The cosmological evolution of the physical characteristics of the cosmological model with parameters
 (\ref{param6}) and initial conditions \eqref{IC9}.
 }
\end{center}
In this case the phase trajectories also adhere to the boundary of the prohibited range however the cosmological acceleration at certain stages is positive and also reaches the values greater than $1$.

%\end{subequations}

%\end{CJK*}
\end{document}